\documentstyle[emulateapj]{article}
\def\eq#1{equation~(\ref{#1})}
\def\Eq#1{Eq.~\ref{#1}}
\def\gsim{{{}_>\atop{}^{{}^\sim}}}
\def\lsim{{{}_<\atop{}^{{}^\sim}}}

\def\kms{{\rm km}\,{\rm s}^{-1}}
\def\kpc{{\rm kpc}}
\def\dls{{D_{\rm LS}}}
\def\te{{t_{\rm E}}}
\def\re{{r_{\rm E}}}
\def\umin{u_{0}}
\def\HJD{{\rm HJD}}
\def\days{{\rm days}}
\def\au{{\rm AU}}
\def\ob14{OGLE-1998-BUL-14}
\def\dos{D_{\rm S}}
\def\dls{D_{\rm LS}}
\def\dol{D_{\rm L}}
\def\thetae{\theta_{\rm E}}
\def\dchit{\Delta\chi^2_{\rm thresh}}
\def\dchif{\Delta\chi^2_{\rm flag}}
\def\dchi{\Delta\chi^2}
\def\elz{{\epsilon}_{\rm LZ}} 
\def\vperp{v}
\def\vtilde{\tilde v}
\def\fb{F_{\rm B}}
\def\fs{F_{\rm S}}
\def\rhos{\rho_{*}}
\def\evmi{{\rm E}(V-I)}
\def\ai{A_{I}}
\begin{document}

\title{Limits on Stellar and Planetary Companions in Microlensing Event OGLE-1998-BUL-14}

\author{
M. D. Albrow\altaffilmark{1}, 
J.-P. Beaulieu\altaffilmark{2},
J. A. R. Caldwell\altaffilmark{3}, 
D. L. DePoy\altaffilmark{4}, 
M. Dominik\altaffilmark{5}, 
B. S. Gaudi\altaffilmark{4}, 
A. Gould\altaffilmark{4}, 
J. Greenhill\altaffilmark{6}, 
K. Hill\altaffilmark{6},
S. Kane\altaffilmark{6}, 
R. Martin\altaffilmark{8},  
J. Menzies\altaffilmark{3}, 
R. M. Naber\altaffilmark{5}, 
R. W. Pogge\altaffilmark{4},
K. R. Pollard\altaffilmark{1}, 
P. D. Sackett\altaffilmark{5}, 
K. C. Sahu\altaffilmark{7}, 
P. Vermaak\altaffilmark{3},  
R. Watson\altaffilmark{6}, 
A. Williams\altaffilmark{8}
}
\author{The PLANET Collaboration}

\altaffiltext{1}{Univ. of Canterbury, Dept. of Physics \& Astronomy, 
Private Bag 4800, Christchurch, New Zealand}
\altaffiltext{2}{Institut d'Astrophysique de Paris, INSU CNRS, 98bis
Boulevard Arago, 75014 Paris, France}
\altaffiltext{3}{South African Astronomical Observatory, P.O. Box 9, 
Observatory 7935, South Africa}
\altaffiltext{4}{Ohio State University, Department of Astronomy, Columbus, 
OH 43210, U.S.A.}
\altaffiltext{5}{Kapteyn Astronomical Institute, Postbus 800, 
9700 AV Groningen, The Netherlands}
\altaffiltext{6}{Univ. of Tasmania, Physics Dept., G.P.O. 252C, 
Hobart, Tasmania~~7001, Australia}
\altaffiltext{7}{Space Telescope Science Institute, 3700 San Martin Drive, 
Baltimore, MD. 21218~~U.S.A.}
\altaffiltext{8}{Perth Observatory, Walnut Road, Bickley, Perth~~6076, Australia}

\begin{abstract}

We present the PLANET photometric data set for \ob14, 
a high magnification ($A_{\rm max}\sim 16$) event alerted by the OGLE
collaboration toward the Galactic bulge in 1998.  
The PLANET data set consists a total of $461$ $I-$band and $139$ $V-$band
points, the majority of which was taken over a three month period.
The median sampling interval during this period is about 1 hour, and the
$1\sigma$ scatter over the peak of the event is $1.5\%$.  
The excellent data quality and high maximum magnification of this
event make it a prime candidate 
to search for the short duration, low amplitude perturbations that
are signatures of a planetary companion orbiting the primary lens.  
The observed light curve for \ob14 is consistent with a
single lens (no companion) within photometric uncertainties.  
We calculate the detection efficiency of
the light curve to lensing companions as a function of the
mass ratio and angular separation of the two components.  
We find that companions of mass ratio $\ge 0.01$ are ruled out at the
95\% confidence level for projected separations between
$0.4-2.4 \re$, where $\re$ is the Einstein ring radius of the primary
lens.  Assuming that the primary is a G-dwarf with
$\re\sim3~{\rm AU}$ our detection efficiency for this event is $\sim 60\%$ 
for a companion with the mass and separation of Jupiter and $\sim5\%$ 
for a companion with the mass and separation of Saturn.  Our
efficiencies for planets like those around Upsilon And and 14 Her are
$> 75\%$.

\end{abstract}

  
\section{Introduction}

Mao \& Paczy{\'n}ski (1991) first suggested that planets could be
detected in microlensing events; Gould \& Loeb (1992) 
pointed out that if all stars had Jupiter-mass 
planets with separations near $5~\au$,
then $\sim 20\%$ of the microlensing events should exhibit detectable planetary deviations,
provided that events were monitored frequently and with
moderately high precision.
Current microlensing discovery teams 
do not generally sample frequently or precisely enough to detect the short-lived perturbations 
caused by planets.  However, since these collaborations reduce their data in
real time, they are able to issue `alerts,' notification of ongoing microlensing events detected 
before the peak magnification.  Prompted by this alert capability,
several other groups have formed to monitor  alerted
microlensing events more closely (GMAN, Alcock et al.\ 1997;
PLANET, Albrow et al.\ 1998; MPS, Rhie et al.\ 1999a).   Since only a handful of
alerted events are in progress at any given time, they can be monitored with
the fine temporal sampling and photometric precision required to 
discover planetary perturbations. In particular, the
PLANET (Probing Lensing Anomalies NETwork) collaboration 
has access to four telescopes that are roughly equally-spaced in
longitude, and thus can monitor microlensing events almost continuously, weather
permitting.  

Over the last five years, PLANET has monitored over 100 events 
with varying degrees of sampling and photometric
precision.  Here we present photometry and analysis of one
such event, \ob14, the 14th event alerted by the OGLE collaboration toward
the Galactic bulge in 1998.  The total PLANET data set for this
event consists of 600 data points, the majority of which was taken during a 3 month
period starting 1 May 1998.  The median sampling interval during this
interval is about $1~{\rm hour}$, with no gaps greater than 4 days.  The
photometric precision near the peak of the event, where the
sensitivity to planets is highest, is $1.5\%$.  These characteristics,
combined with the high maximum magnification of \ob14 make
our data set highly sensitive to planetary perturbations.  The PLANET
data for the event are consistent
with a generic point-source point-lens (PSPL) model.  The excellent
photometry and dense sampling also allow us to 
place stringent constraints on
possible stellar and planetary companions. 
We also place
limits on parallax effects arising from the motion of the Earth,
deviations arising from the finite size of the source, and the amount
of blended light from the lens itself.  These limits are then
translated into limits on the mass of the lens.   We find that,
despite the excellent coverage and photometry of \ob14, the limits on
the mass of the lens are very weak.  This indicates that it will in
general be quite difficult to obtain interesting constraints on the
masses of the lenses giving rise to microlensing events from
photometric data alone. 

Our study is similar to that done by the MPS and MOA
collaborations on the microlensing event MACHO-1998-BLG-35
(\cite{mps}), which was a higher maximum
magnification event ($A_{\rm max} \sim 75$) than \ob14.  We compare
the limits on companions for \ob14 to those for MACHO-1998-BLG-35 directly in \S\ 5.2. 

A brief introduction to the theory of microlensing is given in \S\ 2.
The observations and data are presented in \S\ 3.  In \S\ 4, we discuss known
systematic effects in crowded-field photometry and our method of
correcting for them, and fit the data to a PSPL model.  
In \S\ 5, we search for the kinds of deviations from the PSPL model
that would arise from companions to the primary lens.  Finding none,
we calculate the detection efficiency of
OGLE-1998-BUL-14 light curve to companions as a function of the mass
ratio and angular separation of the
companion, and use this efficiency to place limits on possible
companions to the primary lens.  In \S\ 6, we use several considerations to
constrain the mass of the primary lens.  In \S\ 7, we convert from mass ratio and angular
separation to mass and physical separation of the companion using an
assumption of the mass and distance to the primary lens, and
compare the \ob14 detection efficiencies to other methods of detecting
extrasolar planets.  We conclude in \S\ 8.

\section{Basic Microlensing}

The flux of a microlensing event is given by
\begin{equation}
F(t) = \fs A(t) + \fb,
\label{eqn:flux}
\end{equation}
where $\fs$ is the unlensed flux of the source, $A(t)$ is the
magnification as a function of time, and $\fb$ is the flux of unresolved
stars not being lensed, which may include light from the lens itself.
For a PSPL model, the
magnification is
\begin{equation}
A[u(t)]= { { u^2(t) +2 }  \over {u(t) \sqrt{u^2(t)+4}}},
\label{eqn:mag}
\end{equation}
where $u(t)$ is the angular separation of the source and the 
lens in units of the angular Einstein radius $\thetae$ defined by 
\begin{equation}
\thetae \equiv\left[{4GM \over c^2} {\dls \over \dol \, \dos}\right]^{1/2}
\sim 480 \, \mu{\rm as} \left({M \over M_{\odot}}\right)^{1/2},
\label{eqn:thetae}
\end{equation}
where $M$ is the mass of the lens, and $\dls$, $\dos$, $\dol$ are the
lens-source, observer-source, and observer-lens distances,
respectively.  This corresponds to a physical distance at the lens
plane of
\begin{equation}
\re = \thetae \dol \sim 3~\au \left({M \over 
M_{\odot}}\right)^{1/2}.
\label{eqn:re}
\end{equation}
For the scaling relation on the far right of equations
(\ref{eqn:thetae}) and (\ref{eqn:re}), we have
assumed $\dos=8~\kpc$ and $\dol=6.5~\kpc$, typical distances to the
lens and source for microlensing events toward the bulge.  For rectilinear motion,
\begin{equation}
u(t)= \left[ \left({t-t_0 \over \te}\right)^2 + u^2_{0}\right]^{1/2},
\label{eqn:uoft}
\end{equation}
where $t_0$ is the time of maximum magnification, $\umin$ is the
minimum impact parameter of the event 
in units of $\thetae$, and $\te$ is the Einstein time scale, a characteristic 
time scale of the event defined by
\begin{equation}
\te \equiv { \thetae \dol \over v} \sim 40~\days \left({M\over M_{\odot}}\right)^{1/2}.
\label{eqn:timescale}
\end{equation}
Here $v$ is the transverse velocity of the lens relative to the
observer-source line-of-sight.  For the scaling relation on the far 
right of \eq{eqn:timescale} we have
assumed a transverse velocity of 
$v = 130~\kms$, and we have again assumed $\dol=6.5~\kpc$ and $\dos=8~\kpc$.

A PSPL fit to an observed data set is a function of
$3+2N_l$ parameters:  $\te$, $\umin$, $t_0$,
and one source flux $\fs$ and one blend flux $\fb$ for 
each of $N_l$ light curves taken in different sites or in different bands. 

\begin{figure*}[t]
\epsscale{1.5}
\plotone{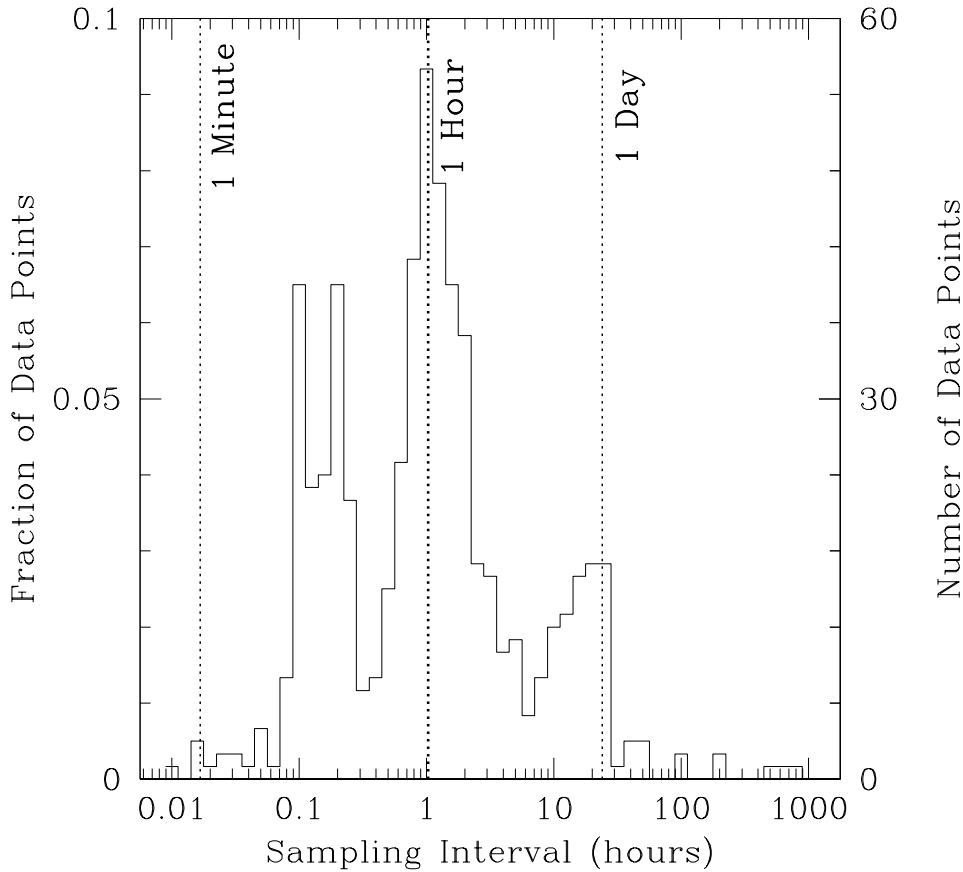}
\caption{
\footnotesize
Histogram of the distribution of sampling intervals (time between successive
measurements) in hours for the \ob14 dataset.  The median sampling interval is
about 1 hour.
}
\label{fig:sampint}
\end{figure*}

\begin{figure*}[t]
\epsscale{1.5}
\plotone{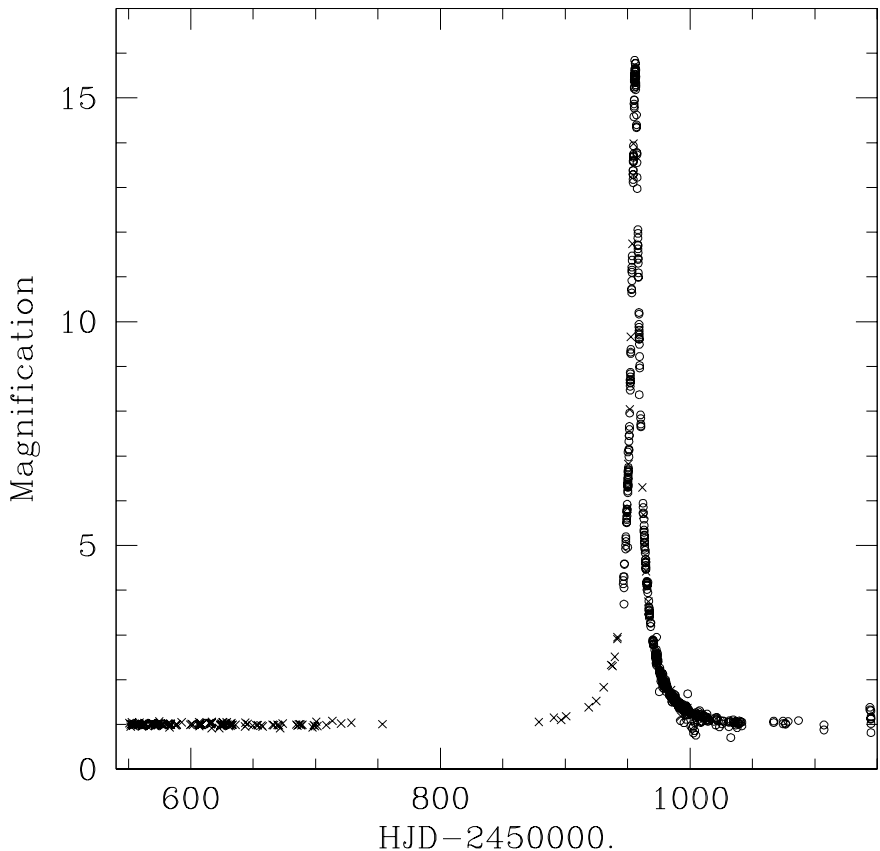}
\caption{
\footnotesize
The combined PLANET (open circles) and OGLE (crosses) light curves for OGLE-1998-BUL-14.  In
order to show all data on the same figure, we have plotted the 
total magnification for each light curve, which is given by $A=(F-\fb)/\fs$, where F is the
observed flux, and $\fs$ and $\fb$ are the source and blended fluxes
derived by fitting the data to a PSPL model, respectively. 
All data prior to $\HJD'\equiv\HJD-2450000 \simeq 940$ are from OGLE;
the majority of the data after $\HJD'=940$ are from PLANET.}
\label{fig:vdata}
\end{figure*}

\begin{figure*}[t]
\epsscale{1.5}
\plotone{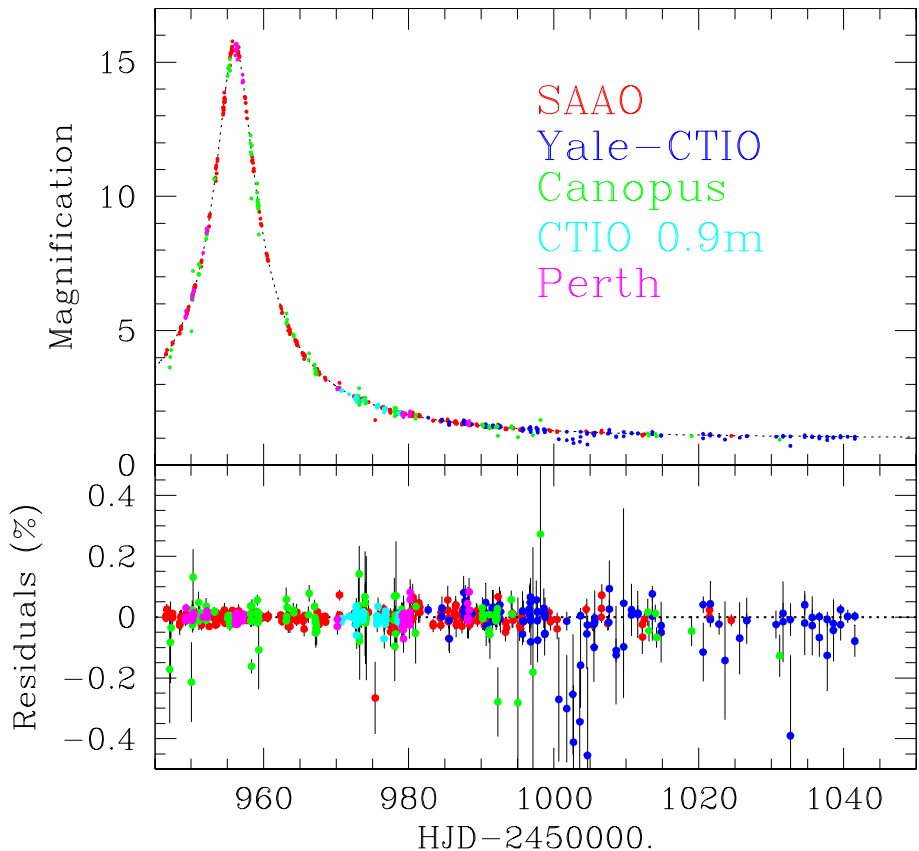}
\caption{
\footnotesize
Top Panel: The cleaned PLANET data for OGLE-1998-BUL-14.  As in Figure~1, the
magnification $A=(F-\fb)/\fs$ is plotted.  Only PLANET data prior to
$\HJD'=1050$, constituting $\sim95\%$ of the complete PLANET
data set are shown.  The dashed line indicates the best-fit PSPL model, 
which has a time scale
$\te=40$ days, and a minimum impact parameter 
$\umin=0.064$, corresponding to a maximum magnification of 
$A_{\rm max}\sim 16$.  The entire data set consists
of 461 $I$-band and 139 $V$-band data points, the majority of which was
taken between $\HJD'\simeq 950$ to
$\HJD'\simeq1040$. The median sampling interval during this time span
is about $1$ hour, or $10^{-3} \te$, with no gaps greater than four days. The data are from the
Yale-CTIO 1m, South African Astronomical
Observatory 1m, the Perth 0.6m, the Canopus 1m, and the CTIO 0.9m. 
Bottom Panel:  Residuals from the best-fit PSPL model.
}
\label{fig:idata}
\end{figure*}

\begin{figure*}[t]
\epsscale{1.5}
\plotone{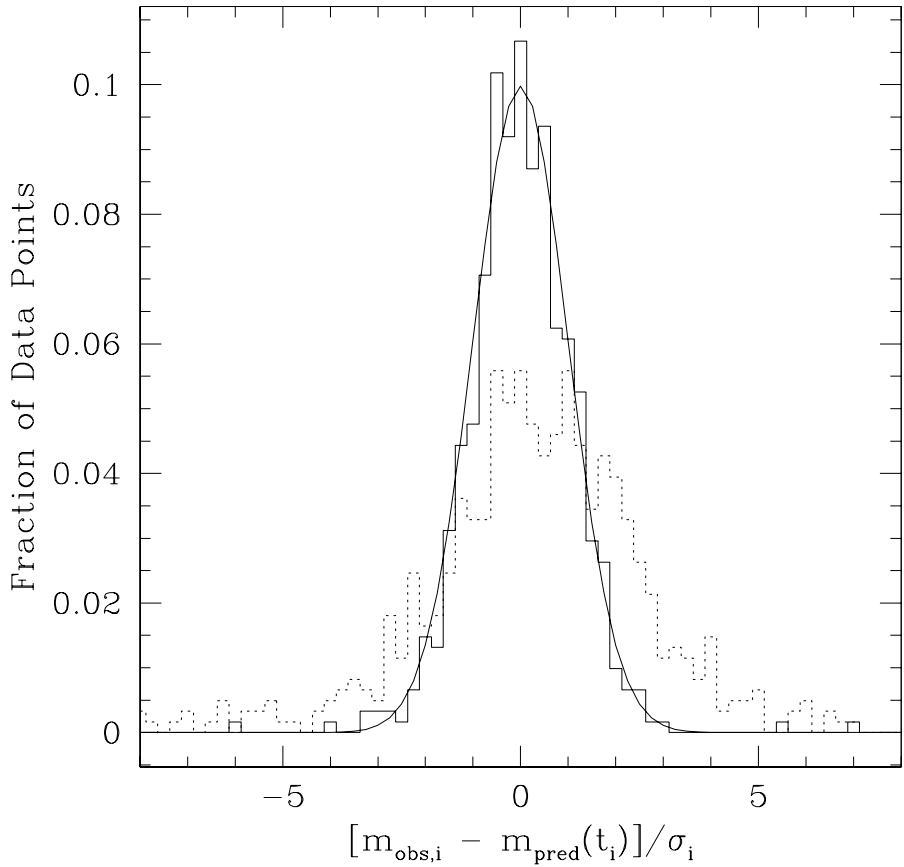}
\caption{
\footnotesize
The histograms show the distributions of the residuals of the
individual data points from the best-fit PSPL model, divided by their
respective errors, before correction for seeing and background
systematics (dashed histogram) and after correction (solid
histogram).  The solid curve is a Gaussian of unit variance.
}
\label{fig:sigmas}
\end{figure*}

\begin{figure*}[t]
\epsscale{1.5}
\plotone{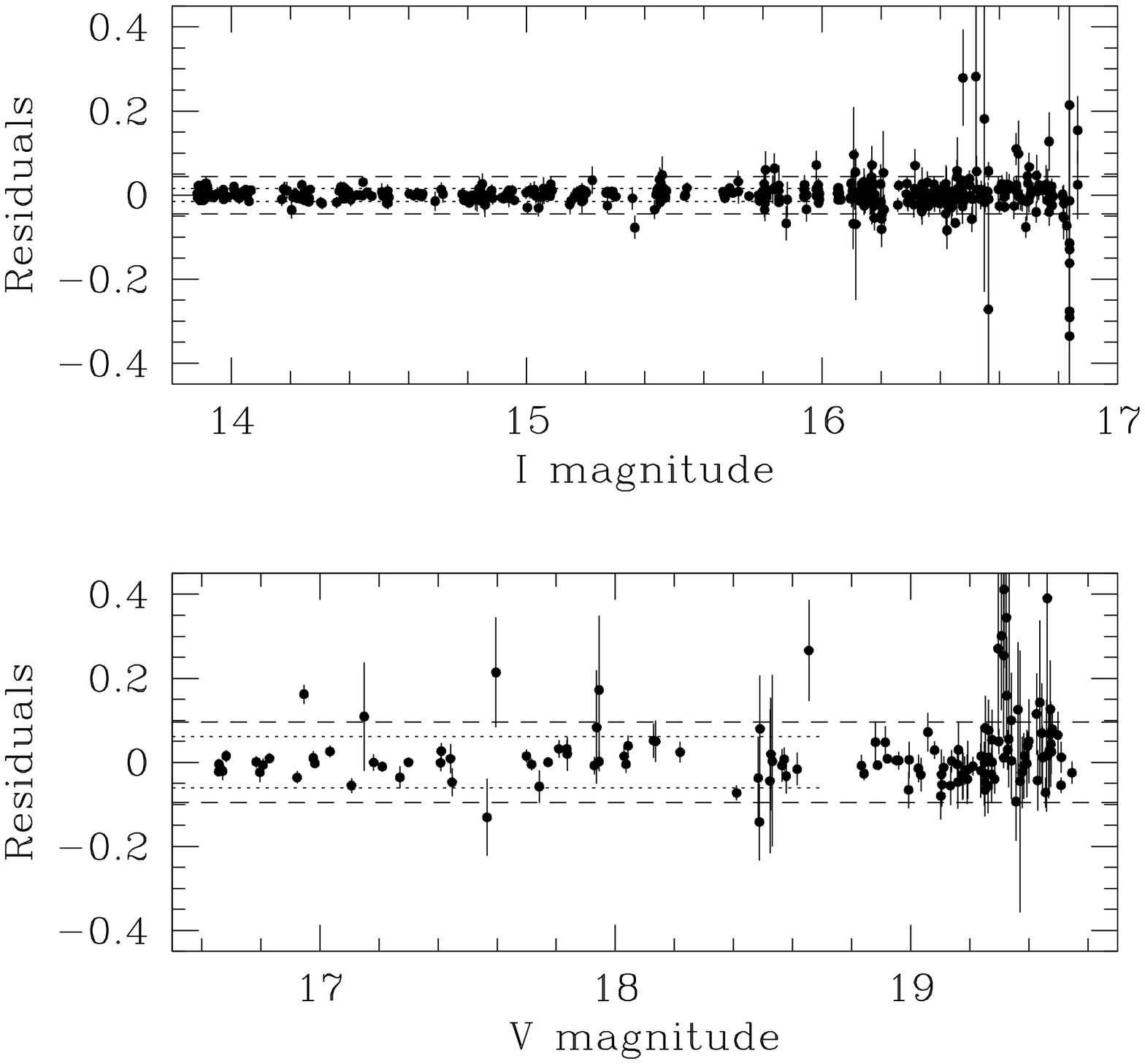}
\caption{
\footnotesize
The points with error bars are the residuals (in magnitudes) of the \ob14 data points
from the best-fit PSPL model as a function of the magnitude predicted
by the model.  We show the residuals separately for the $I$-band
dataset (top panel) and the $V$-band dataset (bottom panel).  
The 1$\sigma$ scatter for the entire $I$-band data set, as denoted by
the dashed line, is $\sim 4\%$, whereas the scatter for all points
with $I < 16$  is $1.5\%$ (dotted line).  The 1$\sigma$ scatter for
the $V$-band data set is $\sim 10\%$ (entire data set), and $\sim 7\%$ ($V<18.7$). 
}
\label{fig:resids}
\end{figure*}

\section{Observations}

PLANET observations of \ob14 were taken in two 
broad-band filters at five sites using seven different
detectors.   The five sites are the CTIO 0.9m and the Yale-CTIO 1m in Chile,
the SAAO 1m in South Africa, the Canopus 1m in Tasmania, and the Perth
0.6m near Perth, Australia.  Canopus data  
prior to $\HJD'\equiv \HJD-2450000.0=975.0$ were taken with a different detector than those taken
afterwards; we will refer to these data sets as Canopus A
($\HJD'<975.0$) and Canopus B ($\HJD'>975.0$), respectively.  SAAO data were taken
in three segments: during
the period from $\HJD'=976.0$ to $\HJD'=980.0$, a different detector
was used than prior to $\HJD'=976.0$ or after $\HJD'=980.0$, when the
original detector was reinstalled on the telescope.  Since different detectors have different
characteristics that can affect the photometry, we will treat these as
independent light curves.  In addition, because the SAAO data prior to
$\HJD'=976.0$ and after $\HJD'=980.0$ are substantially offset
photometrically despite being taken with the same
telescope,  detector and filters, these are also treated as 
independent light curves. We refer to these as SAAO A
($\HJD'>976.0$), SAAO B ($976.0 \le \HJD' \le 980.0$) and SAAO C ($\HJD'
\ge 980.0$). All independent
light curves have both $I$ and $V$ band photometry, except for the CTIO
0.9m and Canopus B data, which do not
have $V$-band photometry.   

The entire data set consists of 461 $I$-band and 139 $V$-band data points forming a total of 14
independent light curves.  The number of data points per light curve is
given in Table 1.  
The data were reduced using DoPHOT (\cite{dophot}), and
reference stars were chosen to optimize photometry at each individual
site.  For details concerning the reduction, see Albrow et al.\ (1998).  

The PLANET data set for \ob14 is one of the best sampled 
light curves to date.  Over $95\%$ of the
measurements were taken during a $3.3~{\rm month}$ time period from
$\HJD'=945.0$ to $\HJD'=1045.0$, corresponding to times $-0.2\te$
before the peak until $2\te$ after the peak.  
In Figure~1, we show the distribution of sampling intervals (time
between successive measurements) in hours
for the cleaned PLANET \ob14 dataset.  The median sampling interval is
about 1 hour, or $10^{-3}$ of the Einstein
ring crossing time.  Furthermore, there are very few gaps greater than
1 day.  

Our primary results are based solely on PLANET data. We
use the publically-available OGLE data
set\footnote[1]{http://www.astrouw.edu.pl/$\sim$ftp/ogle/ogle2/ews/ews.html}
for \ob14 only to test our PSPL model and to derive 
a parallax-based constraint on the lens mass.  The OGLE
data set consists of 159 data points taken in the standard $I$
filter with the 1.3 Warsaw Telescope in Chile: 125 baseline points
were taken prior to $\HJD'=800$ when the source was
not being lensed, and the remaining 34 points taken during the course
of the event.  For more information on the OGLE project and the Early Warning System,
which is used to alert microlensing events toward the Galactic bulge, see
Udalski, Kubiak, \& Szyma{\' n}ski (1997), and Udalski et~al.\ (1994).

\section {Correcting for Known Systematic Effects}
 
Along with the usual reduction procedures, we take
additional steps to optimize the data quality before analyzing the
light curve of \ob14.  Since planetary
perturbations are expected to often have small amplitudes, it is
essential that any low-level systematic deviations caused by
observational effects be minimized in order to avoid spurious
detections.  Such effects can be quite common in crowded-field
photometry. We describe in turn several systematic effects and
our procedures to remove them.

Light curves of constant stars in our fields often display residuals
from the mean value for the star that are correlated with
seeing (or more specifically, image quality as measured by the FWHM of
point sources), and occasionally with the sky background as well.   These
correlations seem to be a generic feature of
crowded-field DoPHOT photometry, and presumably are present at some
level in all light curves.  We find that the magnitude (and sign) of the correlation
depends on the site and detector, being quite strong for some data sets
and almost completely absent in others.  Typically, these
correlations with the seeing and background are linear in flux and
hence approximately linear in magnitudes, and are below 10\%.  
Such correlations inflate the overall scatter in typical 
light curves by about a factor of two, diminishing the recognizability of subtle
deviations.  The correlation with FWHM can be especially
dangerous: coherent deviations of order a few
percent are seen on nightly time scales, partly due to the correlation between
seeing and airmass.  Such deviations can easily mimic low amplitude
perturbations caused by small-mass planetary companions.

Based on studies of constant stars
in crowded fields, Albrow et al.\ (1998) found that formal DoPHOT errors typically
underestimate the true photometric uncertainties by a factor of $1.5-2$.  Using the
formal DoPHOT errors for our analysis would therefore overestimate the significance
of any anomaly being studied.  Furthermore, observed error distributions are
not Gaussian, with long tails toward larger values, 
primarily due to the seeing and background systematics described
above.  These distributions are poorly represented by the formal
DoPHOT errors.  Thus, when not corrected for systematics, many light
curves have more large
($>3\sigma$) outliers than would be expected from a Gaussian
distribution.   While it may be tempting simply to eliminate
these outliers from the data set, such an approach could be dangerous
since an isolated outlier
could, in principle, be due to a short-duration deviation caused by a
planetary companion. Unfortunately, in many
cases the cause of the outliers is not known and nearly-simultaneous
photometry that in principle could be used to
discriminate real deviations from poor-quality data is not available. 

Our approach is to apply a correction to the entire dataset prior to
the analysis, as we now describe.  We fit the
entire data set to a preliminary PSPL model, 
including seeing and background correlation terms, so that for each
observed light curve our model
takes the form,
\begin{equation}
m_{{\rm pred},i}=m_{\rm PSPL} (t_i) + \eta \theta _i + \zeta b_i.
\label{eqn:psplseemodel}
\end{equation}
Here $m_{{\rm pred},i}$ is the predicted flux in magnitudes, $m_{\rm
PSPL}$ is the flux in magnitudes due to the PSPL magnification (\Eq{eqn:flux})
at time $t_i$, $\eta$ is the slope of the seeing correlation, $\theta_i$ is the DoPHOT-reported
full-width half-maximum of the PSF of the $i$th data point, $\zeta$ is
the slope of the correlation with background flux, and $b_i$ is the
DoPHOT-reported background of the $i$th data point.
In general, the specific model used to correct for seeing and 
background systematics is not important,
provided that the model parameters are not strongly correlated with
$\eta$ and $\zeta$.  For most of the deviations we will be considering
in this paper, i.e. those arising from nearly equal mass
binary lenses, parallax, and finite source effects, model parameters
are nearly uncorrelated with $\eta$ and $\zeta$.  However,
deviations caused by small mass ratio companions can occur over the
course of several hours, and thus these deviations can be highly correlated with
variations in seeing.  Since these binary-lens models will not have the benefit of
the additional fit parameters ($\eta,\zeta$), the significance of any short-duration
deviation will be overestimated.  We will return to this point in \S\ 5.2.

We fit the model in the following way.  We choose trial values for
$\te, \umin$, and $t_0$.  This gives a prediction for the
PSPL magnification as a function of time $A(t)$.  The two
parameters $\fs$ and $\fb$ are then determined by performing a linear
fit to the flux.  This gives the PSPL flux $m_{\rm PSPL}
(t_i)$. Finally, the parameters $\eta$ and $\zeta$ are
determined by performing a linear fit in magnitudes.  The final $\chi^2$
for the trial model with parameters ($\te, t_0, \umin, \fs, \fb, \eta,
\zeta$) is then evaluated in magnitudes and the values of these
parameters that minimize $\chi^2$ are then found using a
downhill-simplex method (\cite{cookbook}).  Using the final values of
$\eta$ and $\zeta$, the data are corrected for the seeing and background
systematics.  
This preliminary PSPL fit produces $\chi^2\sim 2000$ for
569 degrees-of-freedom (dof).  Since no gross deviations from a
PSPL light curve are apparent in the data, the high $\chi^2/{\rm dof}$
is an indication that
the DoPHOT errors are underestimating the true error, in this case by
a factor of $\sim 2$.  Some of this inflated
$\chi^2$ arises from a few highly-deviant outliers.  Furthermore, different
light curves have scatter that are underestimated by different amounts.
For these reasons, simply scaling all the errors by a factor that forces
$\chi^2/{\rm dof}$ to be unity would not be appropriate.  We
therefore adopt the following procedure.  Using the preliminary model,
we first find the largest $> 3\sigma$ outlier and reject it.  We
recompute $\chi^2$ for each light curve, and then rescale the errors in
the individual light curves by a factor that forces the
$\chi^2/N_{i}$ for each light curve to be equal to unity, where $N_{i}$ is the number of
measurements in light curve $i$.  All the errors are then scaled with an
overall factor to force  $\chi^2/{\rm dof}$ for the entire data set to 
unity.   This light curve is then refit to the model in
\eq{eqn:psplseemodel}, removing any residual seeing and
background correlations.
We iterate this process, successively removing the largest $3\sigma$ outlier,
rescaling the errors, and refitting the light curve, until no further
$3\sigma$ outliers remain, and the fit has converged.  The final
scaling factors are given in Table 1.  We find a total
of five outliers that deviate from the final model by more than $> 3\sigma$. At the final
step, we form two data sets.  For the first data set, we
reintroduce the $> 3\sigma$ outliers, but with errors scaled by the same
factor as their parent light curves;  we will refer to this as the
``cleaned'' PLANET data set.  The second data set does not contain the
outliers;  we refer to this as the ``super-cleaned'' PLANET data set.
Finally, for some analyses, we combine the OGLE and super-cleaned PLANET data sets,
scaling the OGLE errors by 2.01 so as to force $\chi^2/{\rm
d.o.f.}$ to unity.  We will refer to this as the OGLE+PLANET super-cleaned data set.

We now fit these corrected data sets to a PSPL model with parameters $\te, t_0,
\umin, \fs, \fb$ (\Eq{eqn:flux}).  For the PLANET datasets,
there are 14 light curves, and thus the fit is a function of 31
parameters.  Including OGLE data increases to the number of
parameters by two.  For the cleaned PLANET data set we find 
fit parameters and formal $1\sigma$ errors determined from the
linearized covariance matrix of $\te=(39.6 \pm 1.1)~\days$, $\umin=0.0643\pm 0.0002$,
$t_0=956.016 \pm 0.005$.  The fit parameters for the super-cleaned
PLANET data set and the OGLE+PLANET super-cleaned data set are the same
within the errors.  The parameters and formal $1\sigma$ errors for all three fits are summarized
in  Table 1, along with the blend fractions $g \equiv \fb/\fs$ for
each light curve.  The OGLE+PLANET dataset is shown in Figure~2, while
the cleaned PLANET light curve and the best-fit PSPL model are shown
in Figure~3, along with the residuals from the model.

In Figure~4 we show the distribution of residuals of the cleaned
PLANET dataset from the best-fit PSPL model divided by their
respective (rescaled) errors, along with a Gaussian of unit variance.  Other than
a few outliers, the two distributions are very similar, indicating
that our errors are nearly Gaussian distributed.  In order to
illustrate the importance of including the seeing and background
corrections, we also show the distribution of residuals divided by
their respective rescaled errors before these corrections.  The
uncorrected data have a broader distribution with a median value that
is systematically offset from zero; the
entire distribution is highly non-Gaussian.  

In Figure~5 we show the residuals of the cleaned PLANET dataset from
the PSPL model as a function of the predicted magnitude of the model.
The photometry in $I$ is excellent; the 1$\sigma$ scatter for $I<16$
is 1.5\%; while the scatter for the entire $I$ dataset is $4\%$.  The
scatter is approximately 2.5 times larger in $V$; however this constitutes
only $20\%$ of the entire dataset.

\section{Limits on Companions to the Lens}

The excellent coverage and high-quality photometry, combined with
the fact that \ob14 was a high-magnification event, make 
this an excellent candidate for the detection of planetary
perturbations.  Direct examination of the light curve and residuals
reveals no obvious planetary signatures, and, in fact, no obvious
deviations from the PSPL model of any kind.  However, since the
deviations could be quite subtle, it is important that the light curve
be searched systematically for any deviations.  If no significant
deviations are found, the good photometry and
excellent coverage can be used to place limits on the kinds of companions to the
lens that could be present.  To do this, we must calculate the
detection efficiency of the \ob14 light curve to companions.

\begin{figure*}[t]
\epsscale{2.0}
\plotone{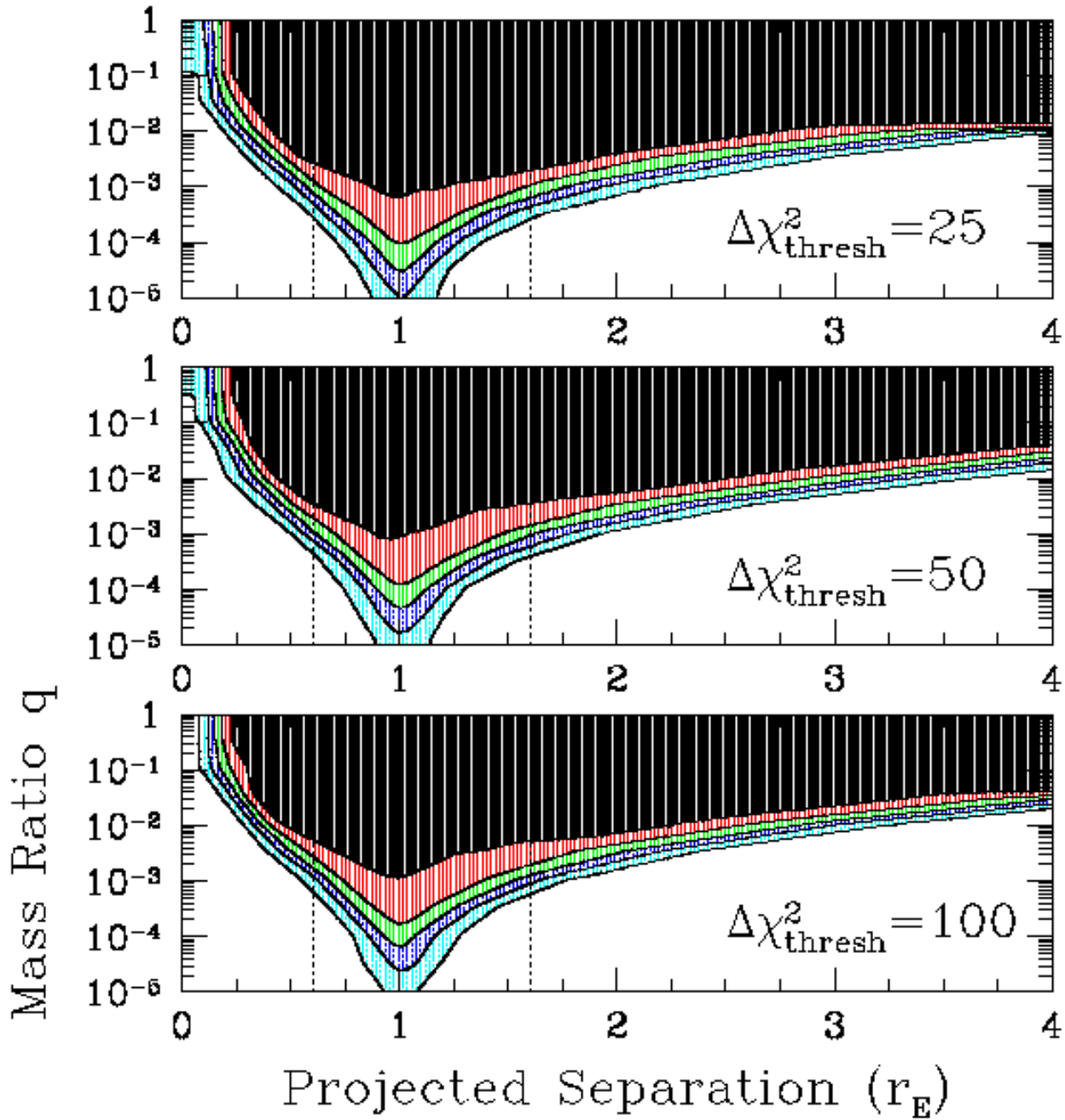}
\caption{
\footnotesize
The solid black lines are contours of constant detection efficiency,
$\epsilon(d,q)$, for all PLANET \ob14 data, shown for projected separations $d$ 
between the primary and companion in units of
the Einstein ring radius, of $0 < d < 4$, and mass ratios
between the primary and companion, $q$, of $0 > \log(q) >-5.0$. 
The contours mark $\epsilon=5\%$ (outer contour), $25\%$, $50\%$, $75\%$,
and $95\%$ (inner contour).  The vertical dotted lines indicate the
boundaries of the lensing zone, $0.6 \le d \le 1.6$.
The region of $(d,q)$ parameter space shaded in black is
excluded by the observations at the $95\%$ confidence level.
The panels are the results for different rejection criteria,
$\dchit=25$ (top), $50$ (middle), $100$ (bottom). 
A mass ratio of $q=10^{-2}$ corresponds to the mass ratio between
a $10 M_J$ planet and a G dwarf. For this
mass ratio, a companion with projected separation $0.4 \le d \le 2.4$ is
excluded at the 95\% confidence level for all our rejection criterion.  
This corresponds to a range of instantaneous projected
separations in physical units of 1.2 to 7.4 AU, assuming a G dwarf primary at 6.5 kpc.}
\label{fig:econtours}
\end{figure*}

\begin{figure*}[t]
\epsscale{1.5}
\plotone{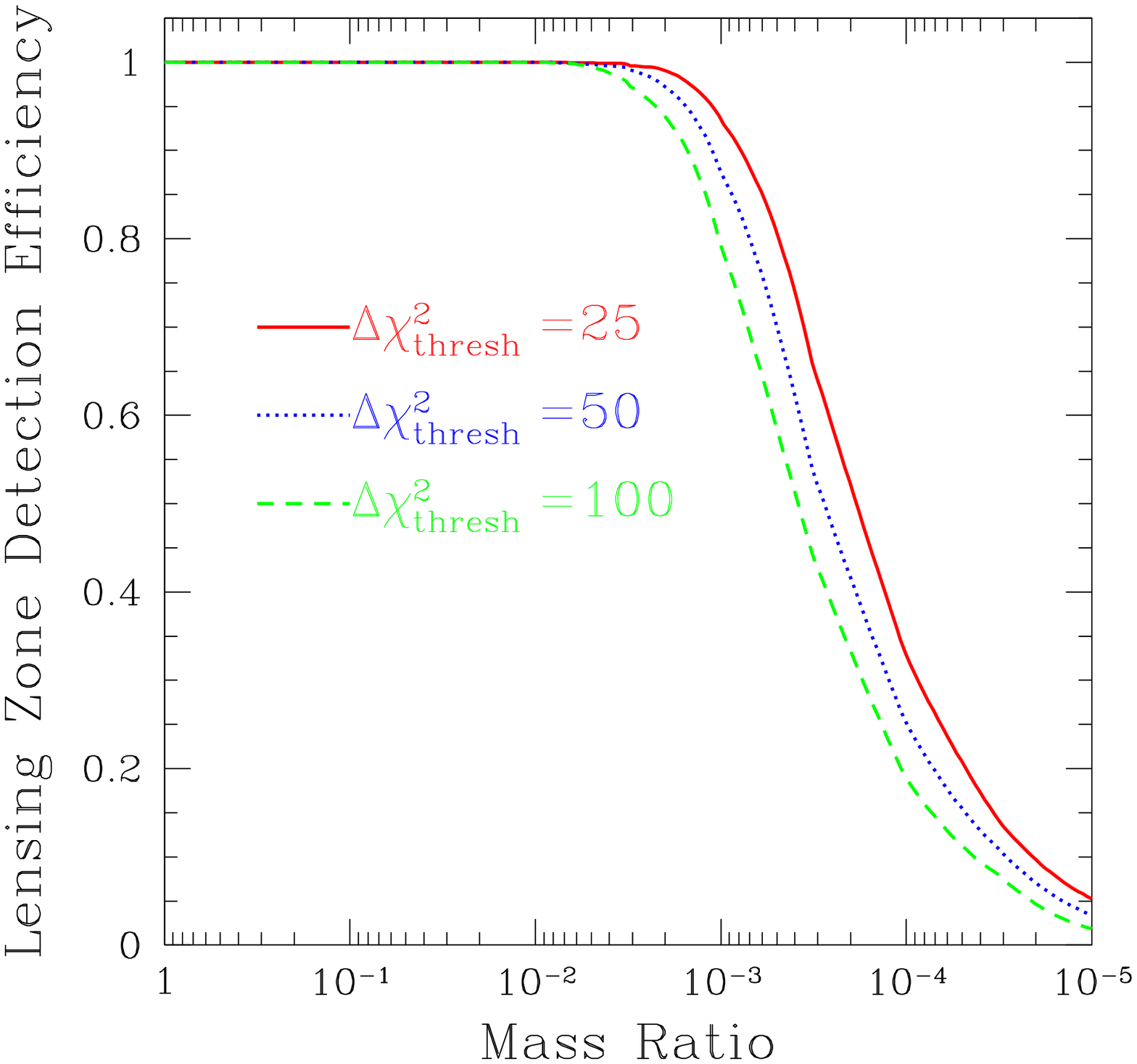}
\caption{
\footnotesize
The detection efficiency of PLANET \ob14 data to a companion averaged over the lensing zone,
$0.6 \le d \le 1.6$, as a function of the mass ratio $q$ between the
primary and the companion.  We show the lensing zone detection
efficiency for three different rejection thresholds, $\dchit=25$
(solid), $50$ (dotted), and $100$ (dashed). 
}
\label{fig:lzeffs}
\end{figure*}

\subsection{Detection Efficiency of \ob14 to Companions}

The efficiency $\epsilon$ of a particular microlensing
light curve to the detection of a binary system 
depends sensitively on two
quantities: the the mass ratio of the system, $q$, and the angular separation $d$
in units of $\thetae$.  The efficiency $\epsilon(d,q)$ 
tends to zero when $q\rightarrow 0$, $d\rightarrow 0$,
or $d\rightarrow \infty$ (i.e. when the companion has a low mass
compared to the primary, or is very close or very far from the
primary).  We simultaneously search for planetary
deviations and calculate the detection efficiency for \ob14 using a
method proposed by Gaudi \& Sackett (2000).  We briefly
review the steps below.
\begin{description}
\item[{(1)}] The \ob14 light curve is fit with a PSPL model by
minimizing $\chi^2$.  The resulting $\chi^2$ for this model is
labelled $\chi^2_{\rm PSPL}$. 
\item[{(2)}] Holding  $d$ and $q$
fixed, the binary lens model
that best fits the observed light curve is found for each source
trajectory $\alpha$, leaving the 31 parameters
($\te, \umin, t_0, 14\times \fs, 14\times \fb$) as free parameters.  
The difference $\dchi(d, q, \alpha)\equiv
\chi^2(d,q,\alpha)-\chi^2_{\rm PSPL}$ is evaluated.
\item[{(3a)}] All parameter combinations ($d,q,\alpha$) yielding
$\dchi(d,q,\alpha) < \dchif$ are flagged for further study as possible
detections,  where $\dchif$ is some
reasonable detection criterion. 
\item[{(3b)}] The fraction of all binary-lens fits for the
given $(d,q)$ that satisfy the criterion $\dchi > \dchit$ in computed, where
$\dchit$ is a rejection criterion.  The
detection efficiency $\epsilon(d,q)$ of the data set for the assumed
separation and mass ratio,
\begin{equation}
\epsilon(d,q)\equiv{1 \over 2\pi} \int_0^{2\pi} {\rm d}\alpha\,\, \Theta[\dchi(d,q,\alpha)
- \dchit],
\label{eqn:ebq}
\end{equation}
where $\Theta[x]$ is a step function, is then computed.  Note that 
$\alpha$ is uniformly distributed. 
\item[{(4)}] Items (2) and (3) are repeated for a grid of $(d,q)$ values.  
This gives the detection efficiency $\epsilon(d,q)$ for \ob14 
as a function of $d$ and $q$,  and also all binary-lens
parameters $(d,q,\alpha)$ that give rise to significantly better fits
to the light curve than the PSPL model.
\item[{(5)}] The parameter combinations found in step (3a) are then
used as initial guesses for our binary-lens $\chi^2$ minimization routine,
leaving all 34 parameters as free parameters, in order to find the
minimum $\chi^2$ and best-fit parameters for this local minimum.
\end{description}

We search for binary-lens fits and calculate $\epsilon(d,q)$ for 
$0 \le d \le 4$ and $0 \ge \log(q)
\ge -5$ at intervals of $0.1$ in $d$ and $0.5$ in $\log(q)$.   
Since the search is performed on a grid of ($d,q,\alpha$) and the grid points are
unlikely to be situated exactly on local minima, it important that
$|\dchif|$
be sufficiently low so that all probable fits are found.  
We chose $\dchif=-9$ for our flagging criterion, and find
only three combinations of $d$ and $q$ for which $\dchi < \dchif$.
Minimizing $\chi^2$ in the neighborhood of these trial solutions
reveals that the best-fit parameters are quite close to the initial
parameters, and $\chi^2$ decreases minimally.  The absolute best-fit binary lens to \ob14 has a $\chi^2= 720.6$, or
$\dchi=-9.2$.  Since this is well below our threshold for detection of
$\dchi=-50$, we do not claim a detection.  Indeed, even a naive 
calculation based on Gaussian statistics and three additional 
parameters to describe the companion yields a chance probability of 
2.7\% for $\dchi\leq -9.2$, which cannot be considered a detection 
at even the 3-sigma level. 

        In fact, the probability of a random fluctuation of this 
magnitude is substantially higher than 2.7\%.  First, since we are including
outlier points in the search, $\chi^2$ should be renormalized by the number
of degrees of freedom, which would imply $\dchi=-7.2$ for which the chance
probability with three additional parameters is 6.7\%.  However, the 
chance probability of a false planet detection is substantially higher
than this.  Moreover, it cannot be computed from a $\chi^2$ table and would
have to be determined by Monte Carlo simulation.  To understand why,
consider one of the planetary models we found that fits \ob14 with 
$\dchi=-7.2$.    The ``success'' of this model is basically driven by
15 points clustered within 1 day whose mean lies $\sim 0.6\%$ above the PSPL 
model.  The chance for such a deviation {\it on this particular day} is
$\sim\exp(\dchi/2)/(2\pi |\dchi|)^{1/2}=0.4\%$.  Since there are a total
of $\sim 600$ data points, there is a similar probability for such a 
fluctuation on each of $600/15=40$ time intervals.  Hence, the total
probability is $1-(1-0.004)^{40}\sim 15\%$.  In fact, the probability is
somewhat higher still because we have taken account of only 15-point
clusters and not larger or smaller clusters that could also mimic
a planet.

        In any event, we have set our detection threshold substantially higher 
than would be warranted solely to avoid chance statistical fluctuations.
This conservative approach is motivated by concern over
unrecognized systematics which, experience has 
taught us, often give rise to spurious detections of formally 
high significance.  We therefore conclude that the light curve of 
\ob14 is consistent with a single lens within the uncertainties.

\subsection{Resulting Constraints on Companions}

The detection efficiency at a given $(d,q)$ is the probability that 
a companion of separation $d$ and mass ratio $q$ would have produced a
deviation inconsistent (in the sense of $\dchi > \dchit$) with the 
observed \ob14 light curve.  If no deviations are seen, companions
with ($d,q$) can be ruled out at a confidence level of $\epsilon(d,q)$.
The choice of the rejection threshold $\dchit$ can have a
significant effect on the resulting detection efficiency, especially
for low thresholds and mass ratios $q\lsim 10^{-3}$
(\cite{meands2000}).  We choose to be conservative, and adopt
$\dchit=50$ as our fiducial threshold.  For comparison, we also show the results for
$\dchit=25$ and $\dchit=100$. 

The binary-lens detection efficiency of \ob14 as a function of the mass ratio $q$ and
angular separation $d$ is shown in Figure~6.  The darkest shading
denotes those parameter combinations $(d,q)$ for which $\epsilon >
95\%$, and thus are excluded at the $95\%$ confidence
level from lying above the rejection criterion. 
Table 2 shows the range of angular separations $d$ that are
excluded by our \ob14 data set for several mass ratios
and the three different rejection thresholds, $\dchit=25, 50$, and $100$.
For $q\gsim 10^{-1.5}$, $\epsilon=100\%$ out to the largest
separation for which we calculate $\epsilon$, namely $d=4.0$.  For
these  mass ratios $d=4$ is thus
a lower limit to the excluded range; the upper end of
the excluded range is likely to be considerably larger. We find that
any companion to
the primary lens with mass ratio $q\gsim
10^{-2}$  and angular separation $0.4 \le d \le 2.4$ is excluded by
the data, that is, such a companion would produce deviations at the
$\dchi > 100$ level that are not observed.  

Note that these limits
apply to individual companions only, not to systems of companions.  
Implicit in our calculation of $\epsilon(d,q)$ is the assumption that
multiple planets affect the magnification structure of the lens, and
therefore the deviation from the single-lens magnification, in an independent way.  This
assumption is likely to break down in regions near the central caustic
when more than one planet is in the lensing zone (\cite{gns}).  In this case,
the efficiencies calculated using the method of Gaudi \& Sackett (2000) will be in error
by an amount that will depend on the relative mass ratios,
orientations, and projected separations of the two companions. Since \ob14
is a high-magnification event, and most of the constraints come from
portions of the light curve near the peak of the event (i.e. where the
central caustic is probed), our results are only strictly valid for single
planets.  A full investigation
of the effect of multiple planets of $\epsilon$ is beyond the scope of
this paper.  We expect, however, that the planet detection
efficiencies for multiple Jovian planetary systems may actually be
higher than for single planets near the peak of \ob14, since the
region of anomalous magnification near the central caustic generally
occupies a larger fraction of the Einstein ring when two planets are
present (\cite{gns}).

It is interesting to compare the limits on companions to the primary
lens of \ob14 to those for the primary lens of MACHO-98-BLG-35 derived by the MPS/MOA collaborations (\cite{mps}).  MACHO-98-BLG-35 was a higher
magnification event $(A_{\rm max}\sim 75)$ than \ob14 $(A_{\rm max} \sim 16)$.
Since higher magnification events have higher intrinsic detection
efficiencies (\cite{gands1998}; \cite{meands2000}), one would expect,
for similar sampling and photometric precision, 
the limits on companions to be more stringent for MACHO-98-BLG-35.
However, although the photometric precision obtained by MPS/MOA on
MACHO-98-BLG-35 is similar to that for \ob14 ($\sim$ few percent), the
sampling of MACHO-98-BLG-35 (in terms of fraction of the time scale
$\te$) is poorer, due primarily to the fact that
MACHO-98-BLG-35 was a shorter time scale event.
Although Rhie et al.\ (1999b) used a slightly different method to
calculate $\epsilon(d,q)$ than that suggested by Gaudi \& Sackett (2000),
and used a different rejection criterion ($\dchit=40$), we can make a
rough comparison between their resulting detection efficiencies shown in their Fig.\
7 with ours for $\dchit=50$ shown in the middle panel of Fig.\ 6.  We
see that the detection efficiencies to companions are everywhere
higher for MACHO-98-BLG-35 than for \ob14.  This indicates that, when
the peak of the event can be measured, maximum magnification is a more 
important factor than sampling in determining
the constraining power (and hence the ability to detect companions) in
an observed microlensing event.

In Figure~7 we show the detection efficiency
averaged over the lensing zone (where
the detection efficiency is the highest), $0.6 \le d \le 1.6$,
\begin{equation}
\elz (q) \equiv \int_{0.6}^{1.6}  \epsilon (d,q)~{\rm d}d~,
\label{eqn:elenszone}
\end{equation} 
for mass ratios $0 \le q \le 10^{-5}$. 
The lensing zone detection efficiencies for several representative
mass ratios are tabulated in Table 3. For a planetary model in which 
companions have angular separations
distributed uniformly throughout the lensing
zone, $\elz$ represents the probability that a companion of mass
ratio $q$ would
have been detected with the \ob14 data set.  These probabilities are
quite high: for example, the detection efficiency for a companion of 
mass ratio $q\gsim 10^{-3}$ in the lensing zone of the \ob14 primary is $\gsim 80 \%$.

For this analysis, we have assumed that the source can be treated
as point-like.  Finite source effects will have a substantial effect on
$\epsilon(d,q)$ if the angular size of source $\rho_*\equiv\theta_* /\thetae$ is comparable to
the Einstein ring radius of the companion, $\theta_{\rm p} = \thetae
q^{1/2}$ (\cite{meands2000}). Any planetary deviations will be
broadened but reduced in amplitude for $\rho_* \gsim q^{1/2}$.   For
\ob14, no finite source size effects were detected, thus we can
place an upper limit on $\rhos$.  As we show in the \S\ 6.2, the $3\sigma$ limit is
$\rhos \le 0.062~$.   Thus the detection efficiencies
calculated above are strictly valid only for $q\gsim 10^{-2.5}$.
However, for typical lens parameters, $\rhos$ is likely to be considerably
smaller, $\rhos\sim0.01$.  Statistically, we thus expect the results to be valid for
$q\gsim 10^{-4}$. For mass ratios less than this, the detection efficiencies calculating
using a point source may be overestimated by tens of
percent (\cite{meands2000}).

Poorly-constrained blend fractions can also induce substantial
uncertainties in the derived detection efficiencies, due to the
correlation between blending and impact parameter $\umin$ (\cite{meands2000}).  Fortunately,
blending is easier to constrain in high-magnification events.  
Indeed, due to the high magnification of \ob14 
and the dense and precise photometry, the blend fraction of the event, and
therefore $\umin$, are quite well
constrained.  The fractional error in $\umin$ is $<1\%$ and 
should contribute negligibly to the
uncertainty in $\epsilon(d,q)$. 

When fitting for the binary-lens model, we did not
include seeing and background correlation terms (\S\ 4).  Although the 
proper approach would be to include these terms in all binary-lens
fits (and indeed, all fits in general), the computational cost is
prohibitive.  Since these terms were included in the PSPL fit, this
implies  that $\dchi$ between the binary-lens fit and single-lens fit 
may be overestimated if the deviations from the PSPL fit due to
the binary lens are highly
correlated with either the seeing or background.  This is generally not a problem
for background correlations because the systematic deviations arising
from the long time scale changes in the
background are unlikely to be confused with deviations caused by
companions.  However seeing correlations
are more insidious, because the short-time scale deviations caused
by low mass ratio companions could be confused with systematic
deviations arising from the nightly
seeing changes in a poorly sampled light curve.  Consider a deviation from the PSPL light curve
that is perfectly correlated with the seeing.  It can be shown that
the fractional error in $\chi^2$ one makes by not including the seeing
correlation in the binary lens fit is $\dchi/\chi^2\sim N_\delta/N_{\rm tot}$,
where $N_\delta$ is the number of deviant data points and $N_{\rm
tot}$ is the total number of data points in the light curve.  For
uniformly sampled data, this is simply the ratio of the time scale of
the perturbation $t_\delta$ to the total duration of observations.  For planetary
microlensing, $t_{\delta}\sim q^{1/2} \te$, and thus
$\dchi/\chi^2\sim q^{1/2}$.  Thus for small mass ratios, $q< 10^{-3}$,
the fractional error in $\chi^2$ is small $\lsim 3\%$.  For $q>10^{-3}$, the 
companions produce coherent deviations that last many days, and thus
cannot be correlated with the seeing.   We therefore conclude that
neglecting the seeing and background correlations does not result in
seriously overestimated detection efficiencies.

\begin{figure*}[t]
\epsscale{1.5}
\plotone{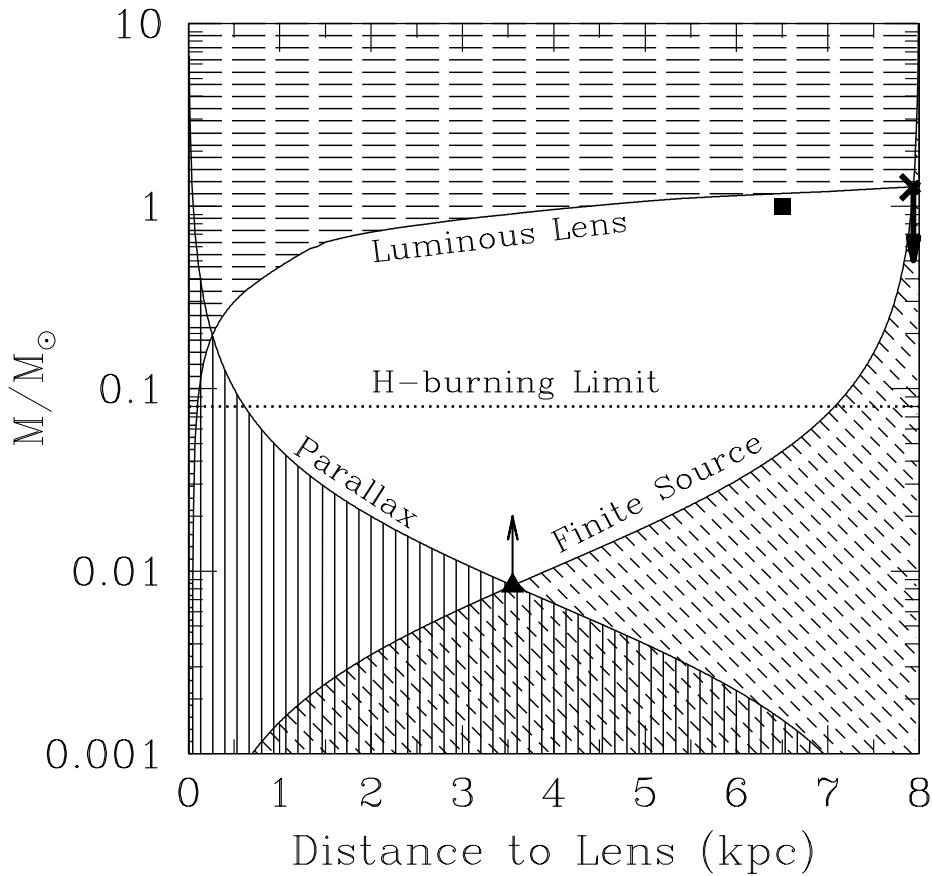}
\caption{
\footnotesize
Excluded regions for the lens mass and distance based on the lack of
second-order signatures in the light curve of \ob14.  
The solid lines show $3\sigma$ upper or lower limits to the mass of
the lens in solar masses $M/M_\odot$ as a function of the distance of
the lens in $\kpc$ for a source distance of $8~\kpc$.  The solid
hatching shows the excluded region from the lack of parallax effects.  The
short-dashed hatching shows the region excluded by the lack of finite
source effects.  Combining the finite source and parallax limits, we
obtain a lower limit to the mass, which is shown as the triangle.  
The long-dashed hatching shows the
region excluded if the lens is a main-sequence star;  the upper limit
to the lens mass is shown as a cross.  The dotted
horizontal line is the hydrogen-burning limit.  Our adopted lens mass
$M=M_\odot$ and distance $\dol=8~\kpc$ are shown as the square.  
}
\label{fig:masslims}
\end{figure*}

\begin{figure*}[t]
\epsscale{1.5}
\plotone{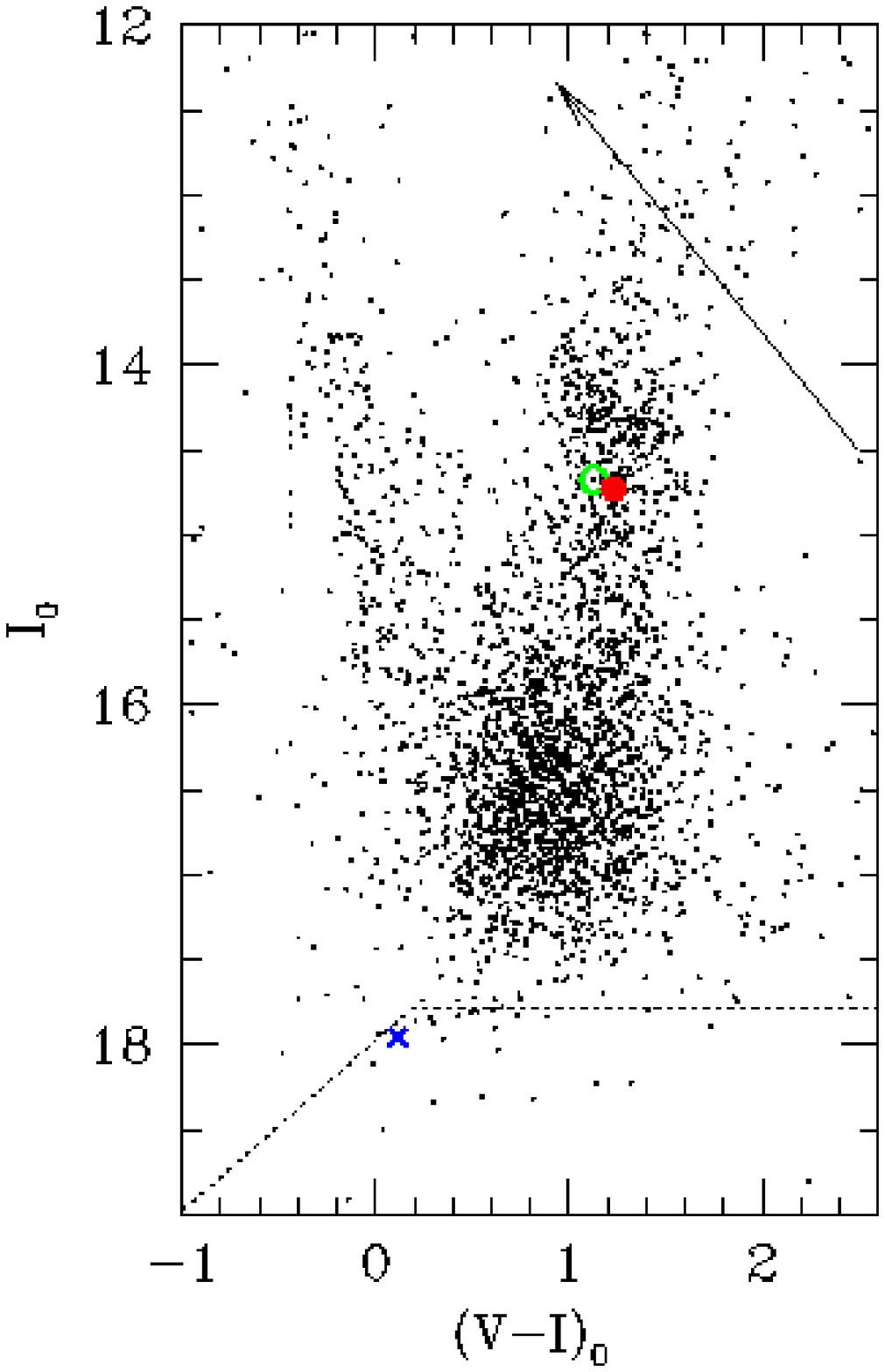}
\caption{
\footnotesize
Calibrated, dereddened color-magnitude diagram for the field centered
on OGLE-1998-BUL-14 from Yale-CTIO 1m data.  The mean reddening for
the field is $A_I$ and E$(V-I)$=1.56, determined from the mean $I$ magnitude and $V-I$
color of the clump.  The arrow shows the direction and
magnitude of the dereddening vector.
The open circle shows the dereddened position of the unmagnified source,
which is composed of the microlensed source (filled circle) and
an unresolved blended star (cross).  The dotted line shows
the $3\sigma$ upper limit to the dereddened I magnitude of the lens,
assuming the extinction to the lens is the same as the \ob14 field.
}
\label{fig:cmd}
\end{figure*}

\section{Limits on the Mass of the Lens}

In the previous section, we placed limits on possible companions
to the primary lens responsible for the microlensing event \ob14 
by calculating the detection efficiency as a function of the mass
ratio $q$\ and  angular separation $d$
between the primary and secondary in units of the angular Einstein
ring of the system.  Although limits on the mass ratio are
interesting in their own right,  what is
of interest ultimately are limits on the mass of the companion,
$M_{\rm p}$,
and the physical orbital separation of the companion, $a$.  In order to
translate limits on $q$ and $d$ into limits on $M_{\rm p}$ and $a$, one must
know the mass of the primary lens, $M$, and its distance $\dol$, and
the orbital phase and inclination of the system.

Unfortunately it is not generally possible to measure the mass and
distance to the lens.  The only parameter
one can measure from a generic single lens event that contains information about
the lens is the time scale $\te$, a degenerate combination of
the mass $M$,  distance $\dol$, and transverse velocity $\vperp$ of
the lens (c.f. \Eq{eqn:timescale}).  However, detection of various
higher order effects, such as parallax or finite source effects,
enables one to extract additional information and partially break this
degeneracy.  In the following subsections, we search for signatures of
these effects in the light curve of \ob14.
Other than a marginal detection of a parallax asymmetry, we do not
detect either of these effects, and use their absence 
to exclude regions of the $(M,\dol)$ plane. In
addition, we place a limit on the amount of light emitted from the lens
itself, and translate this into a limit on the lens mass.  
As we show, despite the excellent photometry and coverage
of \ob14, the limits on these higher order effects are not very
stringent, and do not translate into strong constraints on the mass of the lens.

The analysis uses the super-cleaned PLANET data set since, as
opposed to planetary perturbations, the
effects we are searching for here induce subtle deviations with time scales of many
days, and therefore cannot be described by a isolated
large outliers.  Using the super-cleaned data set (which does not include
these outliers) will provide more robust and reliable limits.  For the
parallax analysis only we use the OGLE+PLANET cleaned data set in
order to have a more secure description of the total baseline flux.

\subsection{Parallax Limits}

The motion of the Earth around the Sun induces departures from
rectilinear motion in the path of the lens relative to the
Earth-source line-of-sight and thus alters the formula for the angular separation
$u(t)$ as a function of time.  This phenomenon is commonly referred
to as parallax, and in general gives rise to light curves that deviate
from the standard PSPL model (\cite{gould1992}; \cite{alcock1995}). 
The magnitude of the deviation depends
on the time of year when the event peaks, the angle $\phi$ of the
lens trajectory with respect to the north ecliptic pole,
and the transverse velocity ${\bf \vtilde} \equiv {\bf \vperp} (\dos/\dls)$ of the lens
projected onto the observer plane.  

If the time scale of the event is short compared to the period of the
Earth's orbit, the Earth's acceleration can be approximated as
constant over the course of the event.  In this case, the angular
separation can be written as (\cite{gbm1994}),
\begin{equation} 
u(t)= \left[ \xi^2 + u^2_{0}\right]^{1/2},~~~~\xi\equiv \tau + {1\over2}\kappa \te \tau^2,
\label{eqn:uoftp}
\end{equation}
where $\tau\equiv (t-t_0)/\te$.  The asymmetry parameter $\kappa$
is given by,
\begin{equation}
\kappa = \Omega_{\oplus} {v_{\oplus}\over \vtilde} \sin\lambda \sin\phi,
\label{eqn:kappa}
\end{equation}
where $\Omega_{\oplus}=2\pi~{\rm yr}^{-1}$, $v_{\oplus}\simeq 30~\kms$
is the linear speed of the Earth around the Sun, and $\lambda$ is the angle between the source
and Sun at the time of maximum magnification.  In the case of \ob14,
$\sin\lambda\sim0.5$. Implicit in \eq{eqn:kappa} is
the approximation that the source is in the Galactic plane.
Thus, for short-time scale events one can measure only the
degenerate combination $\vtilde/\sin\phi$.   

We fit our \ob14 light curve data with the parallax asymmetry model above and
find an improvement of $\dchi=4.4$ for 1 extra dof over the standard PSPL model,
corresponding to a $\sim 2 \sigma$ detection of an asymmetry.  We find
a best-fit value of $\kappa=6.1 \pm 3.6 \times 10^{-4}~(1 \sigma)$,
corresponding to 
\begin{equation}
{\vtilde\over \sin{\phi}}=(420 \pm 250)~\kms~~~(1\sigma).
\label{eqn:vtocos}
\end{equation}
This detection is not particularly useful for our purposes, however,
not only because it is merely a $2\sigma$ detection, but also because
the direction of lens motion is unknown.

In order to obtain a constraint on $\vtilde$ that is independent of
$\phi$, we must use the form for $u(t)$ that includes the full
two-dimensional ($\vtilde,\phi$) parallax information
(\cite{alcock1995}).  Given that the detection of asymmetry was
marginal, it is not surprising that we are unable to obtain
independent constraints on
$\vtilde$ and $\phi$.  
Calculating $\chi^2$ as a function of $\vtilde$, and letting all other
parameters (including $\phi$) vary, we recover the detection of the
parallax asymmetry for $\vtilde \gsim 30~\kms$.  For
$\vtilde\lsim 30~\kms$, however, $\vtilde$ becomes comparable to the Earth's
velocity.  For these small velocities, the component of the Earth's
velocity perpendicular to the motion of the lens becomes significant,
and the light curve begins to deviate appreciably from the PSPL
model, in a manner that is inconsistent with the observations for all
values of $\phi$.  This produces a lower limit to the projected velocity that is independent of
$\phi$,
\begin{equation}
\vtilde > 28.5~\kms~~~~{\rm(3\sigma)}.
\label{eqn:vtbulgemax}
\end{equation}
We can combine this limit with the time scale of the event to give a
lower limit to the lens mass as a function of the relative lens distance,
\begin{equation}
M(x) > 6.7\times 10^{-3} M_{\odot} {{1-x}\over x}~~~~{\rm(Parallax, 3\sigma)}
\label{eqn:plimitbul}
\end{equation}
where $x\equiv\dol/\dos$.  This limit is shown in Figure~8.

For bulge self-lensing, both lens and source belong to a population with
approximately isotropic velocity distributions, producing no preferred
direction for the transverse velocity, $\vperp$, and thus no preferred
value for $\phi$. For disk lenses, however, there is
a preferred direction for $\phi$ due to Galactic rotation, in which
case we would obtain a stronger limit on $\vtilde$.   Unfortunately,
we do not know a priori whether the lens is in the bulge or disk.
We will therefore adopt the conservative assumption that there is no
preferred direction for $\vperp$, and use the limit given in \eq{eqn:plimitbul}.


\subsection{Finite Source Limit}

A point-lens transiting the face of
a source will resolve it, creating a distortion in the
magnification that deviates from the form given in \eq{eqn:mag}.  A detection of
this distortion gives a measurement of the angular size of the source, $\theta_*$,
in units of the angular Einstein ring radius,
$\rhos=\theta_*/\thetae$ (\cite{gould1994}; \cite{neandwick1994};
\cite{wittandmao1994}).  
The requirement for such finite source
effects to be detectable is that the impact parameter of
the event must be comparable to or smaller than the dimensionless source size,
$\umin \lsim \rho_*$.  No such finite-source deviations are
apparent in the light curve of \ob14.  This implies an upper limit to
the dimensionless source size of $\rho_*\lsim0.06$.  For a more exact
limit, we calculate $\chi^2$ as a function of $\rho_*$, leaving all
other parameters free to vary, and find that
\begin{equation}
\rhos < 0.062~~~(3\sigma).
\label{eqn:rholimit}
\end{equation}
In calculating this limit, we have assumed a uniform surface brightness
profile for the source.  Using a more realistic, but model-dependent,
limb-darkened profile weakens this limit slightly.

In order to convert this upper limit on $\rho_*$ into a lower limit on
$\thetae$, we must know the angular size of the source, $\theta_*$,
which can be estimated from its $(V-I)$ color and $I$ magnitude.  Figure~9
shows the calibrated dereddened color magnitude diagram (CMD) for
\ob14 field from Yale-CTIO 1m data.  
The stars in the \ob14  field were calibrated relative to observations
of 9 standard stars in Landolt (1992),
measured several times during the night of 14 August 1998
at the Yale-CTIO 1m telescope in the same $V$ and $I$ filters used
for the observations of the microlensing event. Extinction and
color correction terms were derived in the normal manner. Comparison
with similar measurements on other nights suggests the accuracy
of the calibration is $\sim 0.02$ mag.   We fit the distribution 
of $I$ magnitudes and $(V-I)$
colors of the observed clump to the model of Stanek (1995).  
We then determined the $\evmi$ and $\ai$ by comparing our fitted clump
magnitude and color to the dereddened $I_{\rm cl,0}$ and
$(V-I)_{\rm cl,0}$ of the clump for the bulge as determined by
Paczy\'nski \& Stanek (1998).  We find
$\evmi=1.56$ and $\ai=2.16$.   Using the PSPL fit and Yale-CTIO
standards, we determine the
calibrated, dereddened color and magnitude of the microlensed source 
to be $I_0=14.73\pm 0.03$ and
$(V-I)_0=1.23\pm0.08$, where the errors reflect the calibration
and model uncertainties added in quadrature.   From its position on
the CMD, Figure~9, we conclude that \ob14 is likely to be a clump
giant or a RGB star.

To obtain an estimate of the source radius, $\theta_*$,  we use 
the empirical color-surface brightness relationship for giant stars derived
by van Belle (1999), which we rewrite as
\begin{equation}
\theta_* = 6.0~\mu{\rm as}~10^{-0.2 (V-16) +0.5[(V-I)-1.2]},  
\label{eqn:vbelle}
\end{equation}
where we have assumed the relationship $(V-K)=2.2(V-I)$, derived from
Bertelli et al.\ (1994) isochrones.  For our parameters, we find
$\theta_*=6.3 \mu{\rm
as}$.  An uncertainty in the extinction $\delta A_{\rm V}$ leads to
an uncertainty in the angular size of $\delta \theta_*/\theta_* \sim
-0.16~\delta A_{\rm V}$, where the coefficient depends on the
temperature of the source (\cite{albrowsmc}), which we have assumed
to be $T=5000~{\rm K}$.  We estimate the uncertainty in the extinction
to be $\delta A_{\rm V}\sim0.15$, based on the dispersion of the clump,
resulting in an uncertainty of $\sim 2\%$ in  $\theta_*$.  The $3\%$
uncertainty in the $I$-magnitude of the source leads to an
additional uncertainty of $\sim 1\%$ in $\theta_*$; the $8\%$ uncertainty in the color leads to
an uncertainty of $\sim 6\%$ in $\theta_*$.  Adding these in
quadrature, we find $\theta_* = (6.3\pm 0.4)~\mu{\rm as}$.  

Adopting this value for the angular size of the source, we can now
translate the limit on $\rho_*$ (\Eq{eqn:rholimit}) directly to a limit
on the angular Einstein ring radius,
\begin{equation}
\theta_{\rm E} > 100~\mu{\rm as}~~~~(3\sigma).
\label{eqn:thetaemax}
\end{equation}
Since $\thetae$ depends only on the lens mass and distance 
and on the distance to the source, this limit can be written in terms
of a limit of the lens mass as a function of the relative lens distance,
\begin{equation}
M(x) > 1.0\times 10^{-2} M_{\odot} {x\over{1-x}}~~~~{\rm(Finite~Source, 3\sigma)}
\label{eqn:mulimit}
\end{equation}

We combine the parallax limit (\Eq{eqn:plimitbul})  and the finite source limit
(\Eq{eqn:mulimit}) to obtain a lower limit to the lens mass
$M > 8.3 \times 10^{-3} M_{\odot}$, which occurs at $x=3.55$.  This lower
limit is indicated in Figure~8.

\subsection{Luminous Lens Limit}
If the lens of \ob14 is a
main-sequence star, it emits light.  Although the angular
separation between the lens and source is much too small for the lens
to be resolved, additional light from the lens could, in principle,
alter the shape of the light curve (c.f. \Eq{eqn:flux}).  For most
microlensing events, degeneracies between the fit parameters make it
difficult to constrain accurately the amount of blended light, and
thus any light that may be arising from the lens
itself.  Fortunately, the good photometry, complete
coverage, and small $\umin$ of \ob14 enable the blend fraction to be constrained
quite tightly, allowing us to place an interesting upper limit on the
blended light emitted by the lens.

To do this, we rewrite the flux of any unresolved light as $F_{\rm
B}=F_{\rm U} + F_{\rm L}$, where $F_{\rm L}$ is the flux of the lens,
and $F_{\rm U}$ is the flux of any additional unresolved
source.  Equation \ref{eqn:flux} is then
\begin{equation}
F(t) = \fs A(t) + F_{\rm U} + F_{\rm L}.
\label{eqn:lumlensflux}
\end{equation}
We assume that the
$I$-band flux of the lens is the same for all the $I$-band light curves, and 
similarly that the $V$-band flux is the same for all the $V$-band light curves.
This is a reasonable assumption, since all the observatories use the same $I$ and
$V$ filters, and because the lens and source will be unresolved for all
observatories.  However, due to differences in pixel size and image
quality, we must allow additional unresolved
light, $F_{\rm U}$ not
associated with the lens or source to vary from site to
site.  We then compute $\chi^2$ as a function of $F_{\rm L}$, allowing
the other parameters to vary, and imposing the constraint that $\fb$
always be positive.  For $F_{\rm L} \ll \fs$, $\dchi$ between the fit
with and without a luminous lens is small.  However, as $F_{\rm L}$
increases, $\dchi$ rises because $F_{\rm U}$ must be negative
to match the observations, which is unphysical.  In this way we find
the minimum $I$-band magnitude of the lens that is
consistent with our observations to be
\begin{equation}
I_{\rm L} > \cases{ 21.69 - (V-I) & if $(V-I)\le 1.75$ \cr 
                    19.94         & if $(V-I)\ge 1.75$ \cr},
\label{eqn:imax}
\end{equation}
where the limit is $3\sigma$.  Note that this limit is not dereddened
since the lens may have less extinction than the mean
extinction toward the \ob14 field.  For reference, the dereddened
limit is shown in Figure~9, assuming the reddening to the lens is the
same as for the \ob14 field.

In order to convert this limit on lens light into a limit on the lens mass
as a function of its distance, we must adopt a relationship between
mass and $I$ and $V$ magnitude.  For this purpose, we use the solar metallicity Bertelli
et al.\ (1994) theoretical isochrones for $M>0.6~M_{\odot}$, the solar
metallicity Yale isochrones (\cite{yaleisos}) for $0.35~M_{\odot} \le M \le 0.6~M_{\odot}$,
and extrapolate the Yale isochrones using a 2nd order polynomial for
$0 \le M \le 0.35~M_{\odot}$.  We assume that the dust is distributed
uniformly between the observer and $1.5~\kpc$, with a total reddening at
$1.5~\kpc$ equal to
the mean reddening of the \ob14 field, $\evmi=1.56$ and $\ai=2.16$.
For all distances between 0 and $8~\kpc$, we find the largest mass
that is consistent with the apparent magnitude limit (\Eq{eqn:imax}).
This mass-distance limit is shown in Figure~8.  It can be adequately
represented by,
\begin{equation}
\log[M(x)] < 0.8 +0.07x -0.75 x^{-0.2}~~~~{\rm(Luminous~Lens, 3\sigma)}
\label{eqn:mlumlim}
\end{equation}
Varying the age, metallicity, or dust distribution within reasonable
limits changes the mass limit at any distance by $\lsim 0.2$ dex.

If the lens is a main-sequence star, the largest mass consistent with
the lack of appreciable light in the 
light curve  of \ob14 is $M\sim1.3~M_{\odot}$ which occurs when
$\dol \sim 8~\kpc~(x=1)$.  Thus the lens must be a G dwarf or later.  Of
course, the lens may not be a main sequence star.  
Gould (1999) estimates that $\sim 80\%$ of events detected toward the
bulge are due to main-sequence lenses.   The remaining $20\%$ are due
to stellar remnants, i.e.\ as white dwarfs, neutron stars, or black
holes.  If the lens is such a remnant, it would not obey the
mass-luminosity relationship used to find the limit in
\eq{eqn:mlumlim}.   However, white dwarfs and neutron stars have
masses of $M_{\rm WD}\sim 0.6~M_\odot$ and $M_{\rm NS}\sim
1.35~M_{\odot}$, and so automatically satisfy this limit
for nearly all likely distances to the lens.  Thus the
vast majority (99\%) of lenses will satisfy the limit in \eq{eqn:mlumlim}.

\subsection{Combined Limits}

The limits on the mass and distance to the lens set in the previous
subsections are, for the most part, model independent.  Unfortunately, they are
also not very stringent.  Even with the assumption that the lens is a
main-sequence star, the allowed regions in the $(M,\dol)$ plane
are quite large, spanning two orders of magnitude in mass, $0.01~M_\odot \lsim M
\lsim 1.3~M_\odot$, and nearly the entire range in distance, $0.5~\kpc
\lsim \dol \lsim 8~\kpc$.  Our analysis indicates that, even with
excellent coverage and good photometry, it will be
quite difficult to routinely obtain stringent limits on the mass and
distance to the lens for most events based on photometry alone.

It has been shown by Dominik (1998) how probability densities for
physical quantities of the lens system can be derived under the assumption of
statistical distributions of the mass spectrum, the mass density, and
the transverse velocity.  Rather than doing this, we will simply note that if the lens is
in the bulge ($6~\kpc \lsim \dol \lsim 8~\kpc$),  
and has a typical transverse velocity for bulge
self-lensing events ($v\sim 100~\kms$), then the measured $\te$ implies that it is likely to
have a mass near the upper end of the allowed range.   However, we
cannot rule out that the lens is moving slowly, and therefore that the
mass is quite small.

\section{Discussion}

\subsection{The Detection Efficiency as a Function of Mass and Separation}

In order to convert the limits on companions in the $q-d$ 
plane to limits on companions in the $M_{\rm p}-a$ plane, we need
estimates of the mass and distance to the lens.  However, as we
demonstrated in \S\ 6, it is quite difficult to obtain stringent
limits on these quantities from photometric data alone.  For
illustrative purposes, therefore, we will
simply assume that the lens is a G dwarf and adopt $M=M_\odot$, and a lens distance of
$6.5~\kpc$, so that $\re=3.1$~AU.  We stress, however, that this
choice is somewhat arbitrary, and that the lens
mass may be smaller by two orders of magnitude.

\begin{figure*}[t]
\epsscale{2.0}
\plotone{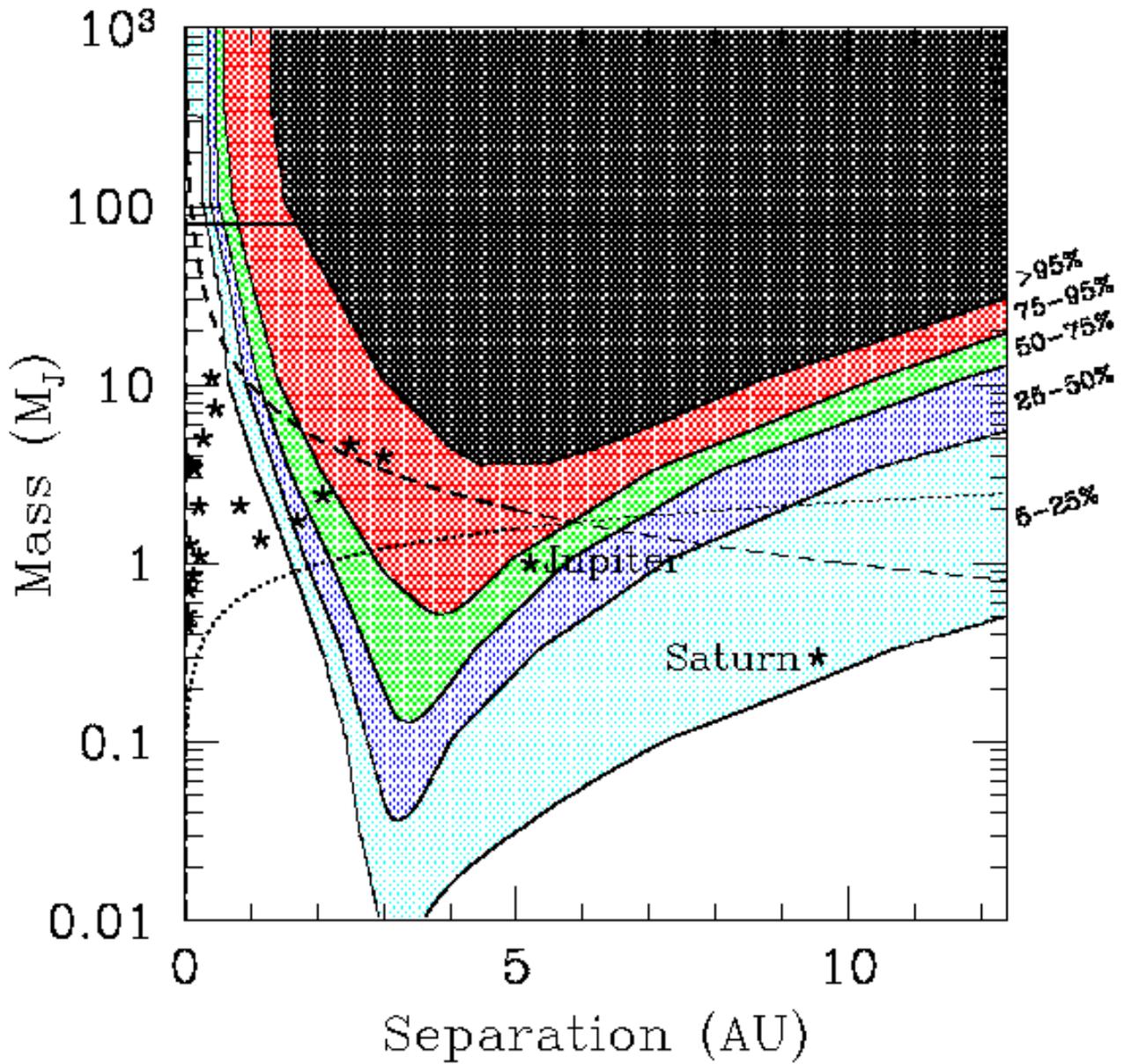}
\caption{
\footnotesize
Example detection efficiencies for the PLANET data set of OGLE-1998-BUL-14 as a
function of the mass and orbital separation of
the companion assuming a primary lens mass of $M_\odot$ and Einstein ring radius of
$\re=3.1~{\rm AU}$.  The contours are the same as in Figure~6. 
In order to convert from mass ratio and projected
separation to mass and physical separation, we have averaged over
orbital phase and inclination (assuming circular orbits).
Jupiter and Saturn are marked with stars, as are
the extrasolar planets discovered with radial velocity techniques.
The horizontal line marks the hydrogen-burning limit.  The dotted line shows the
radial-velocity detection limit for an accuracy of $20~{\rm m s^{-1}}$
and a primary mass of $M_\odot$.  The dashed line is the
astrometric detection limit for an accuracy of $1~{\rm mas}$ and a
primary of mass $M_\odot$ at $10~{\rm pc}$. 
}
\label{fig:mcontours}
\end{figure*}

Since microlensing is only sensitive to the instantaneous angular
separation, $d$, we must first convolve the detection efficiency
$\epsilon(d)$ with the distribution of $d$ for a given semi-major axis
$a$.  To do this we integrate over all random inclinations and orbital
phases, assuming circular orbits.  This distribution is given
explicitly in Gould \& Loeb (1992). Convolving the resulting
distribution with $\epsilon(d)$ gives the detection efficiency as a
function of mass ratio and physical (three dimensional) separation in
units of $\re$.  
We then use the values of $M$ and $\re$ above to convert to
$\epsilon(q,a/\re)$ to the desired detection efficiency
$\epsilon(M_{\rm p},a)$  as a
function of the mass and true orbital separation of the
companion in AU.  

This detection efficiency $\epsilon(M_{\rm p},a)$ as a function of physical
parameters for our assumed primary lens ($M=M_\odot$ and
$\dol=6.5~\kpc$) is shown in Figure~10, for our fiducial rejection threshold of $\dchit=50$.   
Stellar companions to the primary lens of \ob14 with 
separations between $\sim 2~{\rm AU}$ and $11~{\rm AU}$, (the
largest separation for which we calculate $\epsilon$) are excluded. 
 Although we cannot exclude a
Jupiter-mass companion at any separation, we have 
a $\sim 80\%$ chance of detecting such a companion at $3~{\rm AU}$.
The detection efficiency for \ob14 is $>25\%$ at $a=3~{\rm AU}$ for
all companion masses $M_{\rm p}>0.03~M_{\rm J}$.   We find that we had a $\sim 60\%$
chance of detecting a companion with the mass and separation of
Jupiter ($M_{\rm p}=M_{\rm J}$ and $a=5.2~{\rm AU}$),  and a $\sim 5\%$ chance of
detecting a companion with the mass and separation of Saturn ($M_{\rm
p}=0.3~M_{\rm J}$ and $a=9.5~{\rm AU}$) in the light curve of \ob14.

Thus, although Jupiter analogs cannot be ruled out in \ob14, the
detection efficiencies are high enough that future non-detections in
several events with similar quality will be sufficient to place
meaningful constraints on their abundance.

\subsection{Comparison with Other Methods}

How do the \ob14 efficiencies compare to planet detection via other methods?  In Figure~10
we show the radial velocity detection limit on $M_{\rm p}\sin i$ for a
solar mass primary as a
function of the semi-major axis for a velocity amplitude of $K=20~{\rm
m~s^{-1}}$, which is the limit found for the majority of the stars in
the Lick Planet Search (\cite{cummings1999}).  Although we show this
limit for the full range of $a$, in reality the detection sensitivity 
extends only to $a \lsim 5~{\rm AU}$ due to the finite
duration of radial-velocity planet searches and the fact that one
needs to observe a significant fraction of an orbital period.
The limit rises dramatically for  $a\gsim 5~{\rm AU}$ because the period of the companion becomes
larger than the duration of the observations.  In addition, we plot in Figure~10 the $M_{\rm
p} \sin i $ and $a$ for planetary candidates detected in the Lick survey.
Radial velocity searches clearly probe a different region of parameter
space than microlensing, in particular, smaller separations.  Note, however,
that our \ob14 data set gives us a $> 75\%$ chance of detecting analogs to two
of these extrasolar planets: the third companion to Upsilon
And and the companion to 14 Her.  Although the efficiency
is low, we do have sensitivity to planets with masses as small as
$\sim 0.01 M_{\rm J}$, considerably smaller than can be detected via
radial velocity methods.

For comparison, we also show in Figure~10 the astrometric detection
limit on $M_{\rm p}$ for a $M_\odot$ primary at $10~{\rm pc}$,
for an astrometric accuracy of $\sigma_{\rm A}=1~{\rm mas}$.  For an
astrometric campaign of $11$ years, this limit extends to $\sim
5~{\rm AU}$.  Such an astrometric campaign ($\sigma_{\rm A}=1~{\rm mas}, P=11$ years),
would be sensitive to companions similar to those excluded 
in our analysis of \ob14.   The proposed Space Interferometry Mission (SIM) promises $\sim
4~\mu{\rm as}$ astrometric accuracy, which would permit the detection of
considerably smaller mass companions.   

\section{Conclusion}

We have presented the PLANET photometric data set, consisting of 461
$I$-band and 139 $V$-band
measurements, for the microlensing event \ob14.  The median sampling
interval of one hour, RMS scatter of 1.5\% over the peak, and high
magnification of \ob14 make this data set especially sensitive to the
presence of lensing companions.  Within our photometric uncertainties,
the data set is consistent with a single lens.

Our analysis indicates that no companions with mass ratios $q> 0.01$
and instantaneous projected separations $0.4 < d < 2.4$ are
present.  Assuming a solar-mass primary, this mass ratio range includes known stellar binaries and
super Jupiters $(M_{\rm p}=10~M_{\rm J}$). Less massive companions and those at
larger or smaller separations are excluded with less confidence.
Massive companions with $q \gsim 10^{-1.5}$ can be excluded for
projected separations at least as large as $4~\re$. 

The absence of strong parallax, proper motion, or lens light
detections allows us to constrain the mass of the lens to 
$0.01~M_\odot \lsim M \lsim 1.3~M_\odot$.   Assuming a solar-type lens with
$M=M_\odot$ at a distance $\dol=6.5~\kpc$, the Einstein ring radius of
the primary corresponds to $\re=3.1~\au$. Using this value, the PLANET
light curve has efficiencies of 60\% and 5\% for Jupiter and Saturn
analogs, respectively, and a greater than $>75\%$ efficiency for
companions like those in the
Upsilon And and 14 Her systems.  Planets with $M_{\rm p} >10~M_{\rm
J}$ and true orbital separations $1.2~\au < a < 7.4~\au$ are excluded,
assuming these fiducial primary lens parameters.

In performing our analysis, we have considered the systematic effects
of (1) correlations between our photometry and image quality and sky
background; (2) underestimated error bars; (3) finite source size; and
(4) poorly-constrained blending.
We find that the DoPHOT-reported uncertainties
underestimate the true scatter, and that the residuals from the
best-fit model are significantly correlated with
image quality and background.  Applying a simple linear correction
term, rescaling the uncertainties, and eliminating five non-sequential
outliers, however, results in a Gaussian
error distribution yielding $\chi^2/{\rm dof} =1$ for a point-source
point-lens fit.  This allowed us to proceed with a $\chi^2$-analysis
to find the portion of projected separation-mass ratio ($d$-$q$)
binary parameter space excluded by our data.  We estimate that
non-zero source size is unlikely to significantly affect our results
for $q\gsim 10^{-4}$ and almost certainly not for $q\gsim 10^{-2.5}$.
Blending (and thus the true impact parameter) is well constrained for
\ob14 by our data set, and thus has a negligible effect on our
conclusions. 

Our data set for this microlensing event is sensitive to planets
occupying a different range of parameter space than current planet
searches by other techniques.  In particular, super-Jupiters ($M_{\rm
p} \simeq 10~M_{\rm J}$) can be ruled out as companions to the
stellar-mass lens of \ob14 at distances of several AU, larger than
those probed by more sensitive -- but relatively recently commenced --
radial velocity and astrometric searches.

It is not possible to derive general inferences about the abundance
and characteristics of binary or planetary systems from observations
of any single system.   Nevertheless, our results for \ob14 clearly
demonstrate the ability of microlensing to contribute to our knowledge
of Jovian planets several AU from their parent stars, and -- if data
of high enough quality can be collected for a large enough number of
events -- to the search for and study of planets of much smaller mass
as well.  The analysis presented here for \ob14 represents the first
step in the larger task of performing a combined analysis of the
growing PLANET data base of frequently and precisely monitored
microlensing light curves.  When completed, statistical inferences can
be drawn about the frequency and distribution of stellar and Jovian
companions to stellar lenses in the Galaxy.

\begin{acknowledgements}

PLANET thanks the OGLE collaboration for providing real-time alerts and for making
their data publicly available.  We
thank Allison Sills for help with the isochrones.  We are especially
grateful to the observatories that support our science (Canopus, CTIO,
Perth and SAAO) via the generous allocations of time that make this
work possible.  We are indebted to the people that have donated their
time to observe for the PLANET collaboration, including Andreas
Berlind, Alberto Conti, Paul Martini, and Andrew Stephens.  This work was supported in
part by NASA under Award No. NAG5-7589, the NSF under grants 
AST~97-27520 and AST~95-30619, Nederlands Wetenschapelijk Onderzoek
through award GBE 614-21-009, and by Marie Curie Fellowship
number ERBFMBICT972457 from the European Union. 
\end{acknowledgements}

\newpage

\begin{table*}
\begin{center}
\begin{tabular}{|c||c|c|c|}
\tableline
Filter	& Site		& Number of Points & $\sigma/\sigma_{\rm DoP}$\tablenotemark{a} \\
\tableline
\tableline
$I$	& CTIO 0.9m 	&48		& 1.55 \\ 
	& Perth		&50		& 1.00 \\
	& Canopus A  	&60   		& 1.78 \\
	& Canopus B	&47		& 2.25 \\
	& Yale		&56 		& 1.83 \\
	& SAAO A	&113		& 1.43 \\
	& SAAO B	&20		& 1.43 \\
	& SAAO C	&67		& 1.42 \\
\tableline
	& Total PLANET $I$&461		& -- \\
\tableline
\tableline
$V$	& Perth		&4		& 0.08\\
	& Canopus A  	&20 		& 1.77\\ 
	& Yale		&56		& 1.85\\
	& SAAO A	&33		& 1.38\\
	& SAAO B	&6		& 0.94\\
	& SAAO C	&20		& 0.88\\
\tableline
	& Total PLANET $V$&139		& -- \\
\tableline
\tableline
	& Total PLANET $I$+$V$&600 	& -- \\
\tableline
\tableline
$I$	& OGLE		&159		& 2.01\\
\tableline
& Total PLANET+OGLE &754	& --  \\
\tableline
\end{tabular}
\end{center}
\tablenotetext{a}{The scaling factor for the DoPHOT reported errors.}
\tablenum{1}
\caption{ Number of Data Points and Error Scaling Factors \label{tbl:table1}}
\end{table*}
\bigskip

\begin{table*}
\begin{center}
\begin{tabular}{|c||c|c|c|c|}
\tableline
	&		& PLANET CL\tablenotemark{a} & PLANET SC\tablenotemark{b} & OGLE+PL SC\tablenotemark{c} \\
\tableline
\tableline
$t_0$	&		&956.016$\pm$0.005	& 956.011$\pm$0.005
	&956.011$\pm$0.005\\
$\te$	&		&39.6$\pm$1.1		&40.0$\pm$1.2	&40.0$\pm$0.58\\   
$\umin$	&		&0.0643$\pm$0.0002 	& 0.0639$\pm$0.0002&0.0639$\pm$0.0002\\ 
$g_{I}$ & CTIO 0.9m 	&0.16$\pm$0.05		& 0.17$\pm$0.05	&0.17$\pm$0.04\\ 
	& Perth		&0.00$\pm$0.03 		& 0.00$\pm$0.03	&0.00$\pm$0.02\\
	& Canopus A  	&0.03$\pm$0.04   	& 0.04$\pm$0.04	&0.04$\pm$0.04\\
	& Canopus B	&0.14$\pm$0.05		& 0.14$\pm$0.05	&0.14$\pm$0.04\\
	& Yale		&0.03$\pm$0.03  	& 0.05$\pm$0.03	&0.04$\pm$0.02\\
	& SAAO A	&0.18$\pm$0.03		& 0.17$\pm$0.03	&0.17$\pm$0.02\\
	& SAAO B	&0.00$\pm$0.11		& 0.00$\pm$0.11	&0.00$\pm$0.10\\
	& SAAO C	&0.06$\pm$0.03		& 0.06$\pm$0.03 &0.06$\pm$0.03\\
	& OGLE		&--			& --		&0.06$\pm$0.01\\
$g_{V}$  & Perth	&0.14$\pm$0.03		& 0.15$\pm$0.03	&0.15$\pm$0.02\\
	& Canopus A  	&0.47$\pm$0.13  	& 0.20$\pm$0.13	&0.20$\pm$0.13\\ 
	& Yale		&0.12$\pm$0.08   	& 0.13$\pm$0.08 &0.13$\pm$0.07\\
	& SAAO A	&0.30$\pm$0.05		& 0.23$\pm$0.05	&0.23$\pm$0.04\\
	& SAAO B	&0.00$\pm$0.33		& 0.00$\pm$0.33	&0.00$\pm$0.33\\
	& SAAO C	&0.00$\pm$0.07		& 0.00$\pm$0.07&0.00$\pm$0.07\\
$\chi^2$&		&729.7			& 565.3		&720.9     \\
\# points&		&600			& 595		&595+159=754\\
$\chi^2/{\rm d.o.f.}$&	&1.28			& 1.00		&1.00\\
\tableline
\end{tabular}
\end{center}
\tablenotetext{a}{PLANET ``cleaned'' data set; includes all data.}
\tablenotetext{b}{PLANET ``super-cleaned'' data set; does not include
outliers with residuals $\ge 3\sigma$.}
\tablenotetext{c}{Combined PLANET ``super-cleaned'' and OGLE data
sets.}

\tablenum{2}
\caption{ Point-source Point-lens Fit Parameters \label{tbl:table2}}
\end{table*}
\bigskip

\begin{table*}
\begin{center}
\begin{tabular}{|c||c|c|c|}
\tableline
Mass Ratio & $\dchit=25$ & $\dchit=50$ & $\dchit=100$ \\
\tableline
\tableline
$10^{0.0}$   & 0.19--4]\tablenotemark{a}& 0.19--4] & 0.19--4]  \\
$10^{-0.5}$  & 0.19--4]   & 0.19--4]    & 0.19--4]   \\
$10^{-1.0}$  & 0.19--4]   & 0.26--4]    & 0.28--4]   \\
$10^{-1.5}$  & 0.28--4]	  & 0.29--3.86  & 0.29--3.48 \\
$10^{-2.0}$  & 0.38--2.87 & 0.39--2.63  & 0.40--2.37 \\
$10^{-2.5}$  & 0.50--1.96 & 0.57--1.73  & 0.67--1.51 \\
$10^{-3.0}$  & 0.82--1.27 & 0.85--1.20  & 0.88--1.05 \\
\tableline
\end{tabular}
\end{center}
\tablenotetext{a}{Bracket indicates that companions are excluded at
the largest separation we calculate.} 
\tablenum{3}
\caption{Excluded Separations\label{tbl:table3}}
\end{table*}
\bigskip

\begin{table*}
\begin{center}
\begin{tabular}{|c||c|c|c|}
\tableline
Mass Ratio & $\dchi=25$	& $\dchi=50$ & $\dchi=100$ \\
\tableline
\tableline
$10^{0.0}$   & 1.00	& 1.00	     & 1.00\\
$10^{-0.5}$  & 1.00	& 1.00	     & 1.00\\
$10^{-1.0}$  & 1.00	& 1.00	     & 1.00\\
$10^{-1.5}$  & 1.00	& 1.00	     & 1.00\\
$10^{-2.0}$  & 1.00   	& 1.00	     & 1.00\\
$10^{-2.5}$  & 1.00	& 0.99	     & 0.98\\
$10^{-3.0}$  & 0.94	& 0.87	     & 0.80\\
$10^{-3.5}$  & 0.66     & 0.54       & 0.44\\
$10^{-4.0}$  & 0.33     & 0.24       & 0.19\\
$10^{-4.5}$  & 0.14     & 0.11       & 0.08\\
$10^{-5.0}$  & 0.05     & 0.03       & 0.02\\
\tableline
\end{tabular}
\end{center}
\tablenum{4}
\caption{Lensing Zone Detection Efficiencies\label{tbl:table4}}
\end{table*}
\bigskip

\end{document}